\documentclass[fleqn,usenatbib]{mnras}

\usepackage{newtxtext,newtxmath}

\usepackage[T1]{fontenc}

\DeclareRobustCommand{\VAN}[3]{#2}
\let\VANthebibliography\thebibliography
\def\thebibliography{\DeclareRobustCommand{\VAN}[3]{##3}\VANthebibliography}

\usepackage{graphicx}	
\usepackage{amsmath}	

\usepackage{graphics}
\usepackage{hyperref}
\usepackage{multirow}
\usepackage{tablefootnote}
\usepackage{threeparttable}
\usepackage{multirow}
\usepackage{booktabs}

\usepackage{natbib}

\newcommand{\kms}{km~s$^{-1}$}

\newcommand{\msunpc}{$\rm M_\odot~pc^{-2}$}
\newcommand{\msunkpc}{$\rm M_\odot~kpc^{-2}$}

\newcommand{\msunyrkpc}{$\rm M_\odot~yr^{-1}~kpc^{-2}$}
\newcommand{{\hi}}{{H{\sc i}}}
\newcommand{\htwo}{H$_2$}

\newcommand{\gband}{{\em g}-band}
\newcommand{\iband}{{\em i}-band}

\newcommand{\Msun}{M$_\odot$}

\newcommand{\must}{$\mu_\star$}
\newcommand{\nuvr}{$\mathrm{NUV}-r$}
\newcommand{\nuvi}{$\mathrm{NUV}-i$}

\newcommand{\rhi}{$R_{\rm HI}$}
\newcommand{\riso}{$R_{\rm 25}$}

\newcommand{\mycomment}[1]{}

\defcitealias{Lee2025}{L25}

\hypersetup{colorlinks=true,linkcolor=blue,filecolor=blue,urlcolor=blue,citecolor=blue}

\title[\hi\ depletion times within the stellar discs]{WALLABY pilot survey: \hi\ depletion times within the stellar discs of nearby galaxies}

\author[Seona Lee]{Seona Lee,$^{1,2}$\thanks{E-mail: seona.lee@icrar.org}
Barbara Catinella,$^{1,2}$
Tobias Westmeier,$^{1,2}$
Luca Cortese,$^{1}$
Lister Staveley-Smith,$^{1,2}$
\newauthor
Federico Lelli,$^{3}$
O. Ivy Wong,$^{4,1}$
Yago Ascasibar,$^{5,6}$
Alessandro Boselli,$^{7}$
Toby Brown,$^{8,9}$
Nathan Deg,$^{10}$
\newauthor
Akhil Krishna R.,$^{11}$
Denis Leahy,$^{12}$
Syed F. Rahman,$^{13}$
and Jonghwan Rhee,$^{1,4}$
\\
\\
$^{1}$International Centre for Radio Astronomy Research (ICRAR), The University of Western Australia, 35 Stirling Highway, Crawley, WA 6009, Australia\\
$^{2}$ARC Centre of Excellence for All Sky Astrophysics in 3 Dimensions (ASTRO 3D), Australia\\
$^{3}$INAF, Arcetri Astrophysical Observatory, Largo Enrico Fermi 5, 50125, Florence, Italy\\
$^{4}$Australia Telescope National Facility, CSIRO Space \& Astronomy, P.O. Box 1130, Bentley, WA 6102, Australia\\
$^{5}$Departamento de Física Teórica, Universidad Autónoma de Madrid (UAM), Madrid 28049, Spain\\
$^{6}$Centro de Investigación Avanzada en Física Fundamental (CIAFF-UAM), Madrid 28049, Spain\\
$^{7}$Aix Marseille Université , CNRS, CNES, LAM, 13013 Marseille, France\\
$^{8}$National Research Council of Canada, Herzberg Astronomy and Astrophysics Research Centre, 5071 West Saanich Rd. Victoria, BC, V9E 2E7, Canada\\
$^{9}$Department of Physics \& Astronomy, University of Victoria, Finnerty Road, Victoria, BC, V8P 1A1, Canada\\
$^{10}$Department of Physics, Engineering Physics, and Astronomy, Queen’s University, Kingston ON K7L 3N6, Canada\\
$^{11}$Indian Institute of Astrophysics, II Block, Koramangala, Bengaluru 560034, India\\
$^{12}$Department of Physics and Astronomy, University of Calgary, Calgary, AB, T2N 1N4, Canada\\
$^{13}$SBASSE at Lahore University of Management Sciences, LUMS,54792,  Lahore,  Pakistan\\}

\date{Accepted XXX. Received YYY; in original form ZZZ}

\pubyear{\the\year{}}

\begin{document}
\label{firstpage}
\pagerange{\pageref{firstpage}--\pageref{lastpage}}
\maketitle

\begin{abstract}
Neutral atomic hydrogen ({\hi}) reservoirs typically extend far beyond the inner star-forming regions of galaxies, and global \hi\ measurements, which mix these distinct environments, limit our understanding of the gas--star formation cycle.
In particular, global \hi\ depletion times combine gas and star formation from different physical scales, contributing to long measured timescales (5--9 Gyr) and large scatter compared to molecular gas.
Using 841 gas-rich galaxies from the Widefield ASKAP L-band Legacy All-sky Blind Survey (WALLABY) pilot observations, we investigate how \hi\ depletion time and its scaling relations change when \hi\ and star formation are both confined to the stellar disc ({\riso}, the isophotal radius at 25 mag arcsec$^{-2}$ in {\iband}).
We find that depletion times within this region are on average 1.4 Gyr shorter than global values, though some remain very long, indicating that a substantial fraction of \hi\ remains inactive for star formation.
\hi\ depletion times anti-correlate strongly with stellar surface density, and this trend becomes even tighter within the stellar disc.
The Kennicutt–Schmidt relation further reveals an almost constant \hi\ depletion time at fixed stellar surface density, similar to the behaviour seen for molecular gas, suggesting that \hi\ and star formation are regulated by conditions that enable {\hi}--to--{\htwo} conversion, traced by stellar surface density.
Beyond the stellar disc, \hi\ depletion times are on average almost 10 Gyr longer than within {\riso}, confirming extremely inefficient star formation in low-density outer regions.
These results highlight the critical role of spatial location and local conditions for \hi\ to serve as a fuel for star formation.

\end{abstract}

\begin{keywords}
Galaxies: general; galaxies: ISM; galaxies: statistics; radio lines: galaxies
\end{keywords}



\section{Introduction}
\label{sec:int}

Neutral atomic hydrogen ({\hi}) is a key component of the gas--star formation cycle in galaxies, serving as the primary gas reservoir from which dense molecular gas and, ultimately, stars form.
A useful quantity for understanding this cycle is the gas depletion time---inverse of star formation efficiency (SFE)---defined as the timescale over which the gas reservoir would be exhausted at the current star formation rate (SFR), assuming no replenishment of gas from inflows or recycling.
One open question in galaxy evolution is whether \hi\ depletion time provides a physically meaningful link between atomic gas and ongoing star formation.

Global studies suggest that this link is weak.
Typical global \hi\ depletion times ($\mathrm{M_{\rm HI}/SFR}$) are very long \citep[5--9 Gyr, depending on the sample;][]{Huang2012,Wong2016,Saintonge2017,Tudorache2024}, comparable to the age of the Universe and a factor of 2 to 9 longer than depletion times for the molecular gas \citep[$\sim$1--2 Gyr;][]{Saintonge2017,Tacconi2018}, which is more closely related to star formation.
This discrepancy is usually interpreted as evidence that a large fraction of the \hi\ reservoir does not participate directly in star formation, particularly the gas that extends beyond the stellar disc where star formation is significantly less active.

The dependence of \hi\ depletion time on galaxy properties is less well established, with results often varying by survey.
Stellar-mass selected \hi\ surveys like the extended \textit{GALEX} Arecibo SDSS Survey \citep[xGASS;][]{Catinella2018} generally found nearly uniform depletion times with no strong trends and with larger scatter than seen for molecular gas \citep{Schiminovich2010,Saintonge2017}.
In contrast, {\hi}-selected surveys like the Arecibo Legacy Fast ALFA survey \citep[ALFALFA;][]{Giovanelli2005}, Deep Investigation of Neutral Gas Origins \citep[DINGO;][]{Meyer2009}, and the MeerKAT International GigaHertz Tiered Extragalactic Exploration \citep[MIGHTEE;][]{Jarvis2016} reported that galaxies with high stellar mass (above $\sim$10$^{9.5}M_{\rm \odot}$) have slightly shorter depletion times \citep{Huang2012,Jaskot2015,Rhee2023,Tudorache2024}.
The Spitzer Survey of Stellar Structure in Galaxies \citep[$\mathrm{S^4G}$;][]{Sheth2010} also found a similar trend, with starbursts showing shorter depletion times at a fixed stellar mass \citep{Diaz_Garcia2020}.
Stellar surface density often provides a clearer trend: \citet{Jaskot2015} found a strong anti-correlation for non-starburst galaxies, consistent with results from \citet{Wang2017}, while xGASS revealed a weaker but consistent trend with stellar surface density in star-forming main-sequence galaxies \citep{Saintonge2017}. 
In xGASS, the correlation is even stronger with \nuvr\ colour than with stellar surface density for star-forming galaxies \citep{Saintonge2017}.
By contrast, correlations with specific SFR are usually weak or absent \citep[e.g.][]{Jaskot2015,Saintonge2017,Hunt2020}.

These findings have motivated attempts to investigate \hi\ depletion times in the inner regions of galaxies where most star formation occurs.
For instance, \citet{Wang2017} measured \hi\ masses within the optical radius for a sub-sample of the Local Volume \hi\ Survey \citep[LVHIS;][]{Koribalski2018} galaxies, showing that SFE, both globally and within the stellar disc, correlates most strongly with stellar surface density.
\citet{Wang2020} indirectly estimated the \hi\ mass within the optical radius for xGASS galaxies, finding depletion times shorter than global values but longer than 3 Gyr, which are still longer than those of molecular gas.
However, these studies have been limited by modest sample sizes and indirect estimates of \hi\ within the stellar disc, leaving the role of \hi\ in the stellar disc only partially understood.

\hi\ depletion time corresponds to the slope of the Kennicutt-Schmidt (KS) relation between SFR and gas surface densities \citep{Schmidt1959,Kennicutt1998}.
In spatially resolved studies (kpc scales or better), molecular gas shows a tight correlation with SFR and a nearly constant depletion time of 1-2 Gyr \citep[e.g.][]{Bigiel2008,Leroy2013,Pessa2022}, even in the {\hi}-dominated regions \citep{Schruba2011}, underscoring its direct role in fueling star formation.
In contrast, {\hi} correlates only weakly with SFR in inner discs, while in the outer disc the relation is somewhat stronger but highly scattered, corresponding to depletion times of tens to hundreds of Gyr \citep{Bigiel2008,Bigiel2010b,Wang2024}.
Such studies, however, have focused mainly on nearby galaxies with limited samples, so it remains uncertain whether these trends apply across broad galaxy populations.

Recently, \citet[][hereafter, \citetalias{Lee2025}]{Lee2025} directly measured \hi\ and stellar properties within the stellar disc for nearly 1,000 galaxies observed as part of the Widefield ASKAP L-band Legacy All-sky Blind Survey \citep[WALLABY;][]{Koribalski2020} pilot observations \citep{Westmeier2022,Murugeshan2024}, demonstrating stronger links between inner \hi\ reservoirs and star formation activity traced by optical colour.
This result underscores that a critical next step is to understand the regulation of \hi\ depletion time within the stellar disc.
In this study, we extend \citetalias{Lee2025} by incorporating direct star formation measurements to quantify \hi\ depletion times within the stellar disc for the same WALLABY sample. 
By combining {\hi}, stellar, and SFR measurements on matched spatial scales (within {\riso}, the isophotal radius at 25 mag arcsec$\rm ^{-2}$ in {\iband}), we investigate how depletion times change when \hi\ measurements are confined to the stellar disc, and whether this reveals a clearer physical link between \hi\ and star formation than global measurements.

The structure of this paper is as follows. Sections \ref{sec:data}--\ref{sec:method} describe the {\hi}, optical, near-ultraviolet (NUV), and mid-infrared (MIR) data and the measurement of physical quantities. 
Section \ref{sec:sample} outlines the sample selection, Section \ref{sec:result} presents \hi\ depletion time scaling relations and KS analysis, and Section \ref{sec:discus} compares the results to previous studies and discusses them in the context of molecular gas and outer discs, before concluding in Section \ref{sec:conclus}.
This paper uses a flat $\Lambda$CDM model with $H_{0}=$ 70 {\kms} Mpc$^{-1}$ \citep{Riess2016,Abbott2017,Planck2020} and assumes a \citet{Kroupa2002} initial mass function (IMF).

\section{Data}
\label{sec:data}

In this work, we extend the study presented in \citetalias{Lee2025} by incorporating star formation properties.
While we use the same \hi\ and optical data, we also derive star formation properties using data from the Galaxy Evolution Explorer \citep[\textit{GALEX};][]{Martin2005} and the Wide-field Infrared Survey Explorer \citep[\textit{WISE};][]{Wright2010}.

\subsection{\hi\ and optical data}

Here, we briefly summarise the \hi\ and optical data described in \citetalias{Lee2025}.
The \hi\ data are from the first and second Public Data Releases (PDR1 and PDR2) of the WALLABY pilot survey \citep{Westmeier2022,Murugeshan2024}, which provides a suite of \hi\ data products (e.g. source catalogues, spectral line cubes, intensity maps) for over 2,000 \hi\ detections across several targeted fields.
These data offer a spatial resolution of 30 arcsec, a spectral resolution of 4 km s$^{-1}$, and a sensitivity of 1.6 mJy per beam per 4 km s$^{-1}$ channel.
Further details of the observations and data processing are available in \citet{Westmeier2022} and \citet{Murugeshan2024}.

\citetalias{Lee2025} used \hi\ detections from the Hydra cluster, and the NGC 4636, 4808, and 5044 group fields, which contain a total of 1,976 \hi\ detections.
We excluded detections contaminated by nearby radio continuum sources or those that were only partially detected.
For the optical data, we used {\em i}- and \gband\ images from the Dark Energy Camera Legacy Survey (DECaLS), which is part of the Dark Energy Spectroscopic Instrument (DESI) Legacy Survey Data Release 10 \citep{Dey2019}.
We downloaded sky-subtracted {\em i}- and \gband\ cutouts centred on each \hi\ detection and visually inspected them to exclude sources affected by foreground contamination, background artefacts, or ambiguous (i.e. multiple or no) optical counterparts.
This selection resulted in a sample of 1,543 galaxies.

\subsection{\textit{GALEX} and \textit{WISE}}
\label{sec:galex_wise}

For NUV imaging, we retrieved all available \textit{GALEX} NUV tiles ($\lambda_{\rm mean}=2250$ Å) within a circular region centred on each galaxy's \iband\ position, with a diameter matching that of the \hi\ intensity map, to maximise image depth.
We performed this using the astroquery module in Python \citep{Ginsburg2019}.
The downloaded tiles were mosaicked using SWARP \citep{Bertin2002} to produce image cutouts for each galaxy.
Most galaxies were observed in the shallow \textit{GALEX} All-sky Imaging Survey (AIS), and a small number have deeper observations from the Medium Imaging Survey (MIS) or the Nearby Galaxy Survey (NGS).
We visually inspected the final images and excluded those with significant contamination from foreground sources or background artefacts.

For MIR imaging, we used \textit{unWISE} \textit{W}3- and \textit{W}4-bands data \citep[$\lambda_{\rm mean}=12~\upmu m$ and $22~\upmu m$, respectively;][]{Lang2014}, which improved the original \textit{WISE} All-Sky Release by reducing image blurring.
Following the same procedure as \textit{GALEX}, we created cutouts for each galaxy by mosaicking the \textit{unWISE} tiles and visually inspected them to select usable \textit{W}3- or \textit{W}4-band images.
This process resulted in usable NUV and MIR data for $\sim$88\% and $\sim$97\% of the sample, respectively.

\section{Methodology}
\label{sec:method}

\subsection{\hi\ and stellar properties}

\citetalias{Lee2025} derived stellar, global {\hi}, and \hi\ within stellar disc properties, based on the photometry of WALLABY \hi\ intensity maps and DECaLS {\em i}- and \gband\ images.
This includes measurements of stellar isophotal radius at \iband\ surface brightness levels of 25 mag arcsec$^{-2}$ ({\riso}), which we adopt as the stellar disc boundary, the stellar effective radius ($R_{\rm 50\%}$), total stellar mass ($M_{\rm \star}$), and stellar masses enclosed within \riso\ ($M_{\rm \star,R25}$).
In this work, stellar masses from \citetalias{Lee2025} are rescaled from \citet{Chabrier2003} to \citet{Kroupa2002} IMF for consistency with SFR, by multiplying by 1.082 \citep{Madau2014}.
\citetalias{Lee2025} also derived the \hi\ isodensity radius at a surface density level of 1 \msunpc\ ({\rhi}), total \hi\ mass ($M_{\rm HI}$), and \hi\ masses enclosed within \riso\ ($M_{\rm HI,R25}$).
These measurements are further used to calculate average stellar surface density within $R_{\rm 50\%}$ ({\must}$=M_{\rm \star}/(2\pi R_{\rm 50\%}^2)$) and average \hi\ surface densities within {\rhi} and {\riso} ($\Sigma_{\rm HI,RHI}$ and $\Sigma_{\rm HI,R25}$). 
Full details of these derivations are provided in \citetalias{Lee2025}.
The relative mean errors on $M_{\rm \star}$ and $M_{\rm HI}$ for our final sample are $\sigma_{M_{\rm \star}}/M_{\rm \star}=0.16$ and $\sigma_{M_{\rm HI}}/M_{\rm HI}=0.10$.
The error on $M_{\rm \star}$ is estimated from the RMS background noise and the adopted stellar mass-to-light ratio \citep[$\sim$0.10 dex;][]{Taylor2011}, while $\sigma_{M_{\rm HI}}$ is derived from the local RMS noise near the source in the WALLABY source catalogue \citep{Westmeier2022, Murugeshan2024}.

Note that \hi\ masses within the isophotal radii and related properties were measured after degrading DECaLS to WALLABY resolution.
Briefly, we generated a two-dimensional stellar image from the stellar surface brightness profile, convolved it with a 30 arcsec Gaussian kernel representing the WALLABY synthesised beam, and re-measured the isophotal radius from the convolved profile ($R_{\rm 25,c}$).
We did not convolve the actual image because, even with nearby stars masked, faint residual light in the outer regions can spread into the target galaxy after convolution.
$M_{\rm HI,R25}$ was then measured at $R_{\rm 25,c}$ from the \hi\ mass curve-of-growth profile, using elliptical apertures defined by the DECaLS \iband\ centre and position angle, and axis ratios obtained from the \hi\ maps.
Degrading all data to a common 30" resolution provides the most consistent treatment possible with WALLABY, although some methodological biases (e.g. arising from the different radial profiles of \hi\ and stars) are inevitable.
These effects related to beam smearing are likely the main source of systematic uncertainty in our analysis.
Importantly, marginally-resolved and well-resolved galaxies show the same correlations and overall trends, indicating that these systematic offsets are small compared to the underlying relations.

\subsection{Star formation properties}
\label{sec:method_sfr}


We extract the \textit{GALEX} and \textit{WISE} photometry following a similar procedure to that used for DECaLS.
We make a segmentation map to define elliptical apertures and mask sources other than the target galaxy.
The local background is estimated as the sigma-clipped mean image pixel units (ADU) in the annulus between ellipses with semi-major axes of 3 and 5$\times${\riso}.
After subtracting the local background, we calculate the mean ADU in each annulus and convert it to surface brightness using
\begin{equation}
    \frac{m_{\rm \textit{GALEX},NUV}}{\mathrm{mag\, arcsec^{-2}}}=20.08-2.5{\rm \,log\frac{ADU}{P_{\textit{GALEX}}^2}},
\label{eq_ADU}
\end{equation}

\begin{equation}
    \frac{m_{\rm \textit{WISE},\textit{W}3(\textit{W}4)}}{\mathrm{mag\, arcsec^{-2}}}=22.5-2.5{\rm \,log\frac{ADU}{P_{\textit{WISE}}^2}},
\label{eq_ADU}
\end{equation}
where $\mathrm{P_{\textit{GALEX}}}$ (= 1.5" per pixel) and $\mathrm{P_{\textit{WISE}}}$ (= 2.75" per pixel) are the pixel scales \citep{Morrissey2007,Lang2014}.

We estimate the total magnitude in each band from the masked and local background-subtracted images using the asymptotic magnitude method derived from the curve-of-growth \citep[e.g.][]{Munoz-Mateos2015,Reynolds2022}.
However, when the target signal is too weak compared to the noise to derive a reliable surface brightness profile, we instead measure the total magnitude within the \riso\ aperture (4.8\% and 34\% of the sample for \textit{GALEX} and \textit{WISE}, respectively).
\textit{GALEX} NUV magnitudes are corrected for Galactic extinction from the Milky Way using the extinction coefficient of $A(NUV)/E(B-V)=8.2$ \citep{Wyder2007}.
\textit{GALEX} measurements are calibrated on the AB magnitude system, whereas \textit{WISE} magnitudes follow the Vega system. 
Both are converted to luminosity by adopting the local Hubble distance from the WALLABY source catalogue\footnote{Our sample likely includes cluster members. For example, \citet{Reynolds2023} find that 34\% of the \hi\ detections in the Hydra field are cluster galaxies. Nonetheless, the main quantities used in this paper (e.g., {\hi} depletion time, {\hi} surface density, etc.) are independent of distance, meaning that uncertainties in the adopted distances do not affect the trends presented in this study.} \citep{Westmeier2022,Murugeshan2024} as the luminosity distance.

We measure the total SFR by summing the unobscured and obscured SFRs, following \citet{Reynolds2022}:
\begin{equation}
    \mathrm{SFR=SFR_{NUV}+SFR_{\textit{W}3(\textit{W}4)}}.
\end{equation}
The unobscured SFR is estimated from \textit{GALEX} NUV-band luminosity using the calibration from \citet{Schiminovich2007}:
\begin{equation}
    \mathrm{SFR_{\rm NUV}}/(\mathrm{M_{\odot}\, yr^{-1}})=10^{-28.165}{L_{\rm NUV}}/({\mathrm{erg\, s^{-1}\, Hz^{-1}}}).
\label{eq_ADU}
\end{equation}
The obscured SFR is derived from \textit{WISE} mid-infrared luminosities, following \citet{Jarrett2013}, which is widely adopted in previous studies \citep[e.g.][]{Janowiecki2017,Janowiecki2020}:
\begin{equation}
\begin{aligned}
    \mathrm{SFR_{\rm \textit{W}3}}/(\mathrm{M_{\odot}\, yr^{-1}})=4.91\times10^{-10}{L_{\rm \textit{W}3}}/L_{\rm \odot},\\
    \mathrm{SFR_{\rm \textit{W}4}}/(\mathrm{M_{\odot}\, yr^{-1}})=7.50\times10^{-10}{L_{\rm \textit{W}4}}/L_{\rm \odot}.
\end{aligned}
\label{eq_ADU}
\end{equation}

Although alternative calibrations may shift the absolute SFR values (median $\mathrm{SFR_{\rm \textit{W}3,J13}}/\mathrm{SFR_{\rm \textit{W}3,C17}}$ = 3.8, where each obscured SFR is based on \citet{Jarrett2013} and \citet{Cluver2017}, respectively), these systematic offsets are unlikely to affect the observed trends and correlations presented in this work.
The typical uncertainty in total SFRs derived from hybrid \textit{GALEX} and \textit{WISE} measurements is $\sim$0.2~dex \citep{Kennicutt2012}.
We note that there are calibrations specifically designed for such hybrid tracers \citep[e.g.][]{Leroy2019,Belfiore2023}. 
In this work, we adopt the same calibrations as the xGASS survey \citep{Janowiecki2017} to enable direct comparison with their sample in Section \ref{sec:discus1}. 
Using the \citet{Leroy2019} calibration instead would change the total SFR by only a factor of 1.1 (0.04 dex).
We find a systematic decrease in the unobscured SFR with increasing galaxy inclination; however, the dependence on inclination becomes less evident when considering the total (i.e. obscured plus unobscured) SFR.
We confirmed that the galaxy inclination does not influence the key trends in our analysis, and this dependence disappears for quantities measured at 30" resolution.

We use the same approach as for measuring \hi\ mass to estimate the SFR within the stellar disc.
Specifically, we generate two-dimensional images from the surface brightness profiles in the \textit{GALEX} NUV and \textit{WISE} \textit{W}3 and \textit{W}4 bands.
Each image is convolved with a Gaussian kernel whose full width at half maximum (FWHM) is $\sqrt{30^2-\theta^2}$ arcsec, where $\theta$ is the native resolution of the image, to match the 30 arcsec resolution of the \hi\ data.
Unlike the DECaLS \iband\ image ($\theta\approx$ 1"), the resolutions of \textit{GALEX} NUV ($\theta\approx$ 5.6"), \textit{WISE} \textit{W}3 ($\theta\approx$ 6"), and \textit{WISE} \textit{W}4 ($\theta\approx$ 12") are not negligible.
We calculate cumulative NUV (MIR) magnitudes (curve-of-growth) using elliptical apertures defined by the galaxy centre and position angle from the DECaLS i-band image, with axis ratios from the convolved NUV (MIR) image.
The cumulative NUV and MIR magnitudes at the $R_{\rm 25,c}$ radii are then converted into SFR$_{\rm NUV,R25}$ and SFR$_{\rm \textit{W}3(\textit{W}4),R25}$ following the same procedure as for the total magnitudes, and final SFR$_{\rm R25}$ are obtained by combining them.

Although \textit{W}4, which traces hot dust emission from small grains, generally provides a more reliable SFR estimate, and \textit{W}3 can be affected by PAH emission and older stellar populations \citep{Calzetti2007,Engelbracht2008,Leroy2019}, we primarily use \textit{W}3-based SFRs since \textit{W}3 images have higher sensitivity than \textit{W}4 \citep{Wright2010,Lang2014}, resulting in lower noise and fewer artefacts when measuring $\mathrm{SFR_{\textit{W}3,R25}}$.
When \textit{W}3 photometry is unreliable (e.g. due to contamination by foreground sources), we instead use the \textit{W}4-based SFR (<1\% of the sample).
We compared \textit{W}3- and \textit{W}4-based SFRs to assess the impact of older stellar populations and PAH emission on \textit{W}3. The median difference is small ($\mathrm{\log (SFR_{W3}/SFR_{W4})\sim0.08}$) and shows no clear dependence on SFR or stellar mass.
We do not apply corrections for contamination from older stellar populations, as this effect is almost negligible for our gas-rich, star-forming galaxies (median $\mathrm{SFR_{\textit{W}3,corr}/SFR_{\textit{W}3}}$ = 0.99).
Importantly, a substantial portion of the total SFR in our sample is traced by the unobscured component (median fraction = 0.65), further reducing the impact of this uncertainty in the obscured SFR on our results.

Finally, we derive the global \hi\ depletion time ($t_{\rm dep}{\rm (HI)}$) and specific SFR ($\mathrm{sSFR = SFR/M_{\rm \star}}$).
The mean measurement uncertainty in the \hi\ depletion time is 0.2~dex.
The average SFR surface density within the \hi\ disc is derived as
\begin{equation}
    \Sigma_{\rm SFR,RHI}\equiv\frac{\rm SFR}{\pi R_{\rm HI}^2}.
\end{equation}

\begin{figure}
\centering
\includegraphics[width=\linewidth]{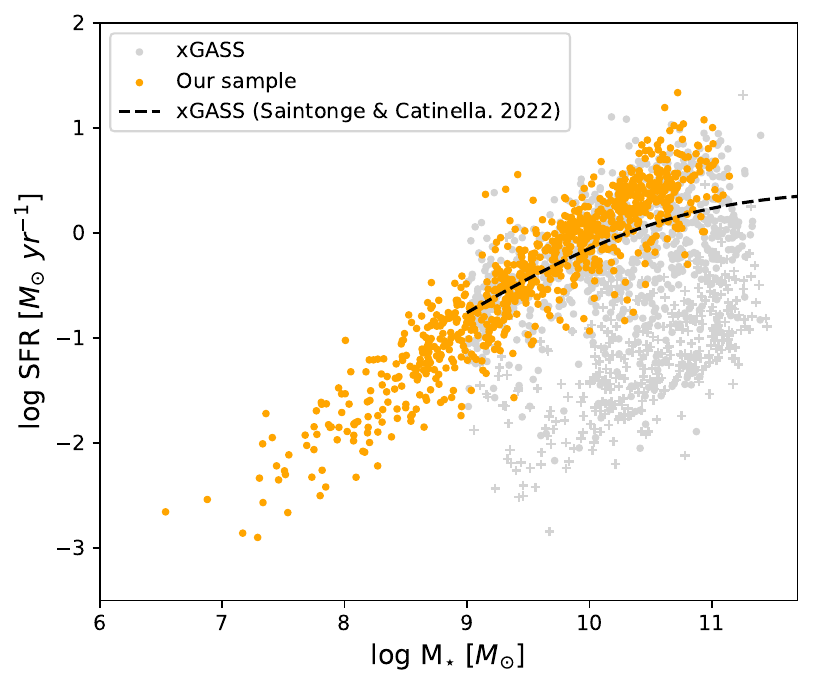}
\caption{Distribution of WALLABY galaxies (orange points) in the SFR versus stellar mass plane. The grey points show xGASS detections (circles) and non-detections (crosses), and the black dashed line indicates the star-forming main sequence from \citet[][based on xGASS]{Saintonge2022}.}
\label{fig_sfms}
\end{figure}

Using the SFR$_{\rm R25}$ measurements, we calculate the \hi\ depletion time and the average SFR surface density within \riso\, and the \hi\ depletion time outside the stellar disc:
\begin{equation}
    t_{\rm dep}{\rm (HI)}_{\rm R25}\equiv\frac{M_{\rm HI,R25}}{{\rm SFR}_{\rm R25}},
\end{equation}
\begin{equation}
    \Sigma_{\rm SFR,R25}\equiv\frac{{\rm SFR}_{\rm R25}}{\pi R_{\rm 25,c}^2},
\end{equation}
\begin{equation}
    t_{\rm dep}{\rm (HI)}_{\rm out}\equiv\frac{M_{\rm HI}-M_{\rm HI,R25}}{{\rm SFR}-{\rm SFR}_{\rm R25}}.
\end{equation}
In cases where the \hi\ radius is smaller than the stellar radius ($R_{\rm HI}<R_{25,c}$; $\sim$10\% of the sample), quantities within the stellar disc are computed within \rhi\ rather than \riso\ to ensure consistency, i.e. measuring {\hi}, stars, and star formation properties on the same spatial scale.

\begin{figure*}
\centering
\includegraphics[width=\linewidth]{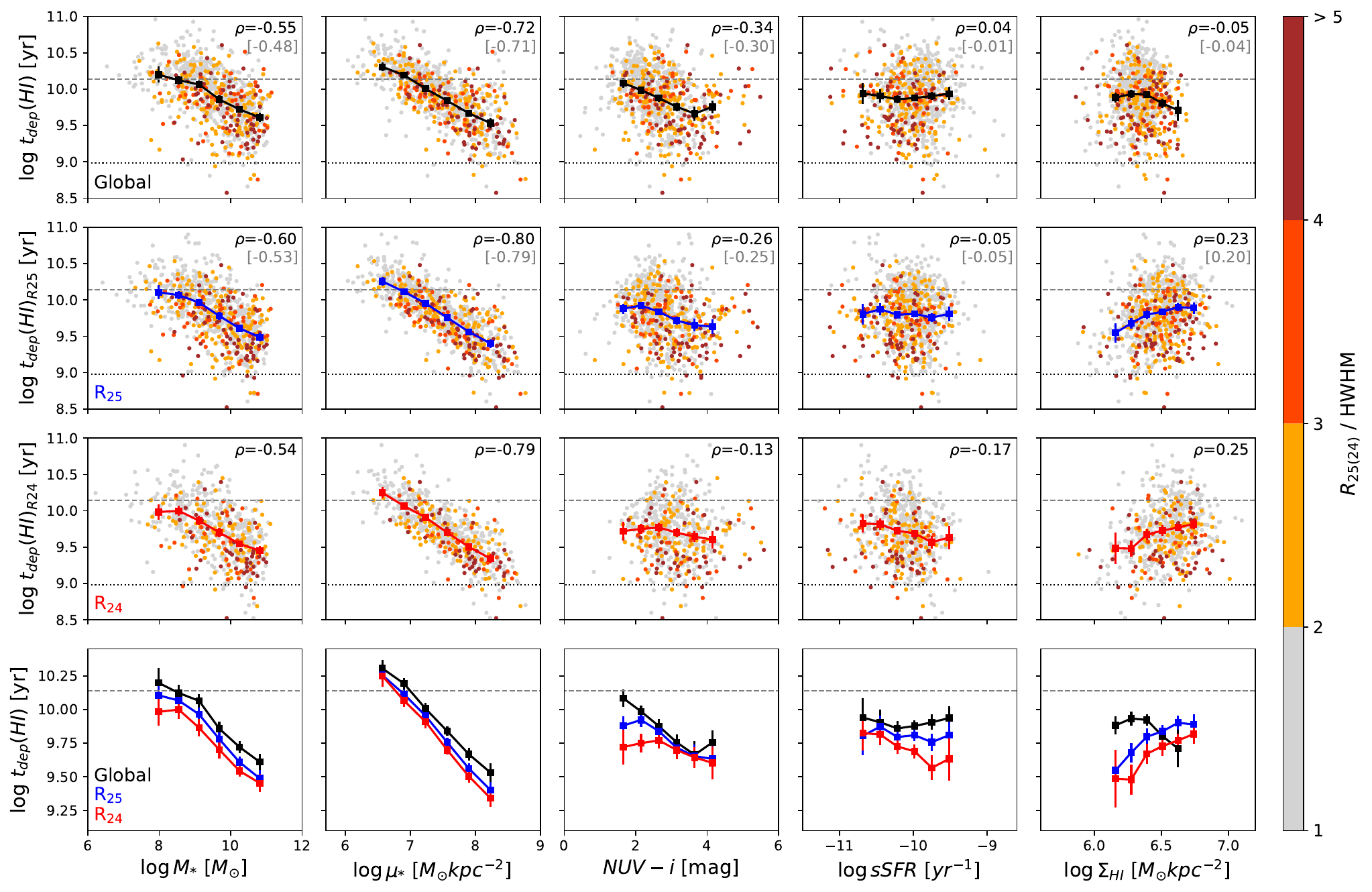}
\caption{Scaling relations of \hi\ depletion time measured globally (first row), within {\riso} (or \rhi\ if $R_{\rm HI}<R_{\rm 25,c}$; second row), and within $R_{\rm 24}$ (third row) as a function of stellar mass, stellar surface density, {\nuvi} colour, sSFR, and average \hi\ surface density (left to right columns).
All quantities are measured within the corresponding spatial scale, except for stellar surface density, which is always averaged within the effective radius.
Squares show the average of log ($t_{\rm dep}$) in each bin containing at least 10 galaxies, with error bars showing the standard error of the mean. These are replotted in the bottom row for comparison. The grey dashed and black dotted horizontal lines indicate the age of the Universe \citep[13.8 Gyr;][]{Planck2020} and mean \htwo\ depletion time for main-sequence galaxies \citep[0.95 Gyr;][]{Saintonge2017}, respectively. Spearman coefficients are shown in the upper right corner, and those for the $R_{\rm 24}$-based sample (third row) are given in brackets for direct comparison. Galaxies are colour-coded by resolution, from grey to darker colours, based on the number of beams across the stellar disc ({\riso}/HWHM for the top two rows; $R_{\rm 24}$/HWHM for the third row), where HWHM is WALLABY's beam's half-width half maximum (=15").}
\label{fig_HI_dep_scaling}
\end{figure*}

\section{Sample selection}
\label{sec:sample}

Our sample builds on that of \citetalias{Lee2025}, who measured \hi\ and stellar properties for 1,543 WALLABY galaxies.
To ensure reliable measurements of \hi\ mass within the stellar disc, \citetalias{Lee2025} selected galaxies with stellar diameters larger than the WALLABY beam FWHM of 30" (i.e., at least one beam across the stellar disc), yielding 995 galaxies with {\riso} > 15".\footnote{The resolution is estimated using the original {\riso} rather than the convolved one ($R_{25,c}$), providing a more physically meaningful measure of how well the galaxy is resolved.}
From these, we further select galaxies with reliable total SFR measurements.
This includes 595 galaxies well detected in NUV and MIR bands and 246 galaxies where one measurement is an upper limit and is estimated from the total magnitude within the \riso\ aperture, although its contribution to the total SFR is almost negligible.
Our final sample therefore consists of 841 galaxies, with a median distance of 97 Mpc, for which the median \riso\ of 24" corresponds to a physical size of 11 kpc.
Of these, 66\% have stellar discs resolved by one to two beams, 21\% by two to three beams, and 6\% by three to four beams.
The \hi\ discs are, on average, 1.9 times larger than the stellar discs in our final sample, and 50\% of the galaxies have \hi\ discs resolved by more than three beams.
We confirm that even when the \hi\ radius is smaller than the stellar radius, the \hi\ disc is still larger than the WALLABY beam, ensuring that all measured quantities are resolved by at least a single beam.

Fig.~\ref{fig_sfms} shows the distribution of the sample in the SFR-stellar mass plane.
Galaxies broadly follow the star-forming main sequence (dashed line), consistent with gas-rich star-forming characteristics of WALLABY galaxies \citep[see also][Fig.~1]{Reynolds2023}.
Compared to xGASS, however, they show a narrower SFR range at fixed stellar mass, i.e. narrower sSFR, likely due to WALLABY's shallower \hi\ sensitivity and the selection of galaxies with \riso\ > 15", which excludes more {\hi}-rich and star-forming systems at higher redshift \citepalias[Fig.~2 in][]{Lee2025}.
Galaxies with stellar mass $\rm >10^{10}M_{\odot}$ tend to lie above the star-forming main sequence, suggesting that WALLABY may not be fully representative of the main-sequence population at those stellar masses.
This limited range of sSFR may reduce apparent correlations presented in the paper.

\section{Results}
\label{sec:result}

Long \hi\ depletion times \citep[5--9 Gyr;][]{Huang2012,Wong2016,Saintonge2017,Tudorache2024} are generally interpreted as evidence that much of the \hi\ reservoir does not directly participate in star formation, particularly the large fraction located beyond the stellar disc where star formation is largely inactive.
Furthermore, previous studies have reported weak or inconsistent correlations between global \hi\ depletion times and galaxy properties, often with substantial scatter, partly due to sample selection, limited statistics, and the inclusion of outer {\hi} \citep[e.g.][]{Huang2012,Saintonge2017,Wang2017,Hunt2020,Tudorache2024}. 
Here, we use the large WALLABY pilot sample to examine \hi\ depletion time scaling relations on both global and stellar disc scales and test whether these trends become stronger when \hi\ is restricted to the stellar disc.

\begin{figure*}
\centering
\includegraphics[width=\linewidth]{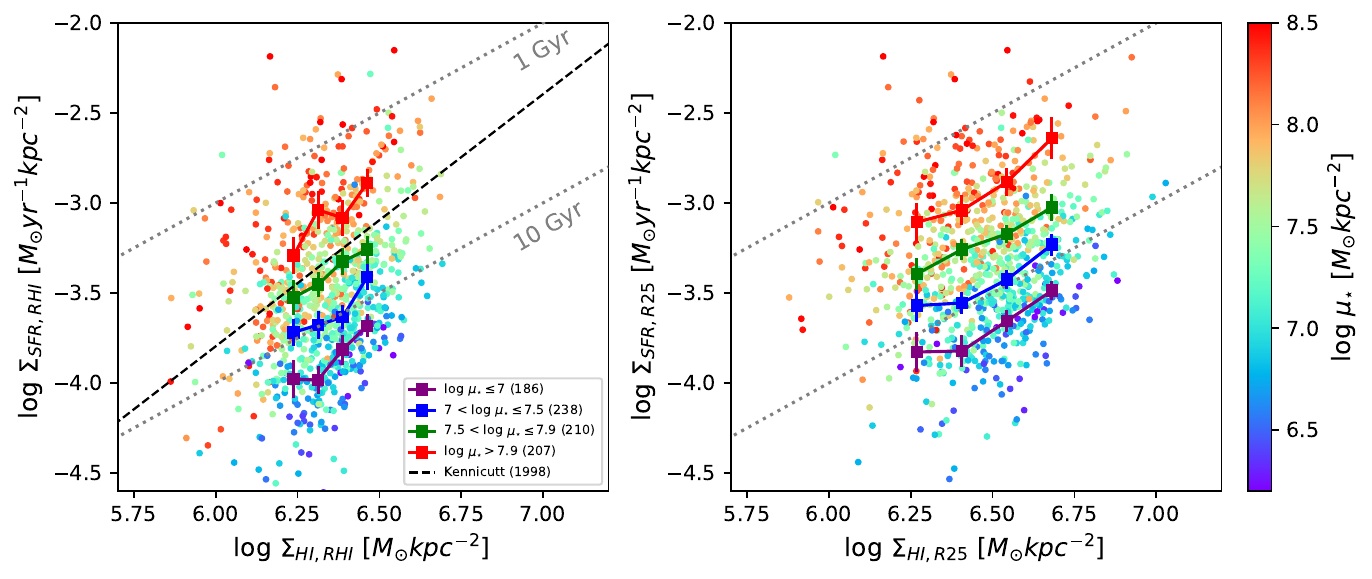}
\caption{Scaling relations between average SFR and \hi\ surface densities measured within \rhi\ (left) and \riso\ (or \rhi\ if $R_{\rm HI}<R_{\rm 25,c}$; right). Galaxies are colour-coded by their stellar surface density. Squares show the average of $\log (\Sigma_{\rm SFR})$ in four stellar surface density bins, with error bars indicating the standard error of the mean. The range and the number of galaxies in each stellar surface density bin are indicated in the lower right corner of the left panel. The grey dotted diagonal lines indicate \hi\ depletion times of 1 and 10 Gyr, and the dashed black line shows the global relation derived by \citet{Kennicutt1998}. Mean measurement uncertainties in $\rm \Sigma_{HI}$, $\rm \Sigma_{SFR}$, and $\rm \mu_{\star}$ are 0.05, 0.2, and 0.07~dex, respectively.}
\label{fig_KS_in}
\end{figure*}

\begin{figure}
\centering
\includegraphics[width=0.95\linewidth]{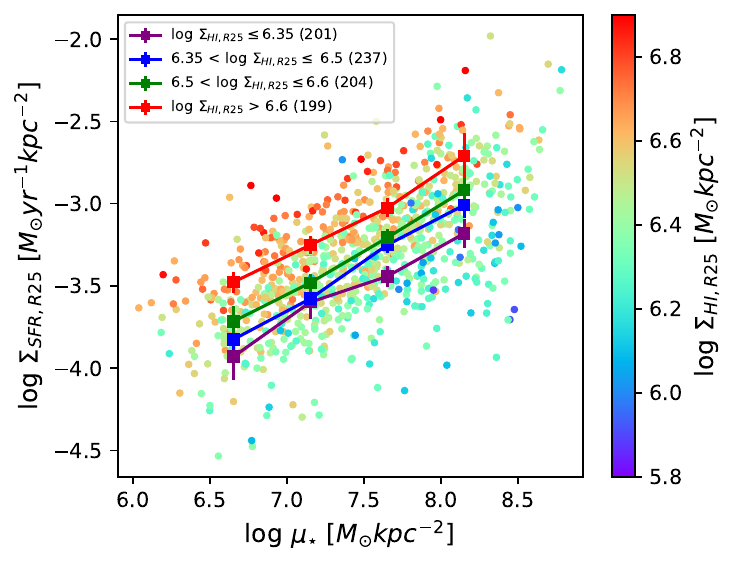}
\caption{Relation between average SFR surface density within {\riso} and stellar surface density colour-coded by average \hi\ surface density within {\riso} (or \rhi\ if $R_{\rm HI}<R_{\rm 25,c}$). Squares show the average of $\log (\Sigma_{\rm SFR})$ in four \hi\ surface density bins, with error bars indicating the standard error of the mean. The range and the number of galaxies in each bin are shown in the upper left.}
\label{fig_mu_star_sfr}
\end{figure}

\begin{figure*}
\centering
\includegraphics[width=0.9\linewidth]{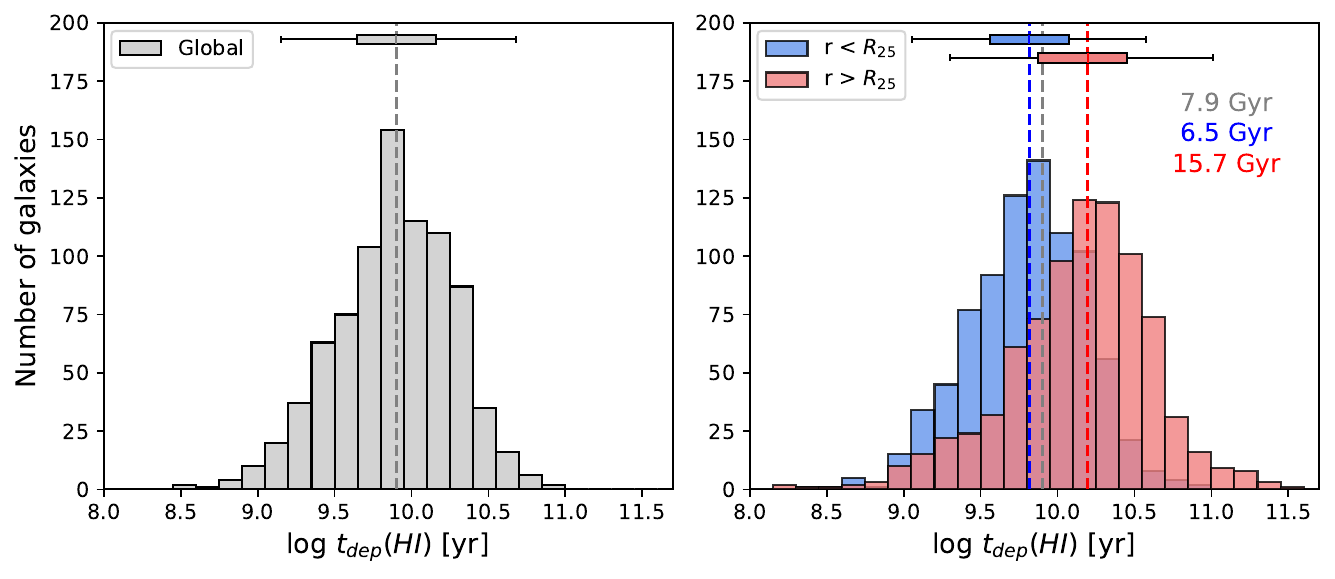}
\caption{Histograms of \hi\ depletion time measured globally (left; black) and within and outside \riso\ (or \rhi\ if $R_{\rm HI}<R_{\rm 25,c}$; right; blue and red, respectively).
Dashed lines indicate median values (shown on the right side of the right panel) and the whisker box plots show the medians and interquartile ranges.}
\label{fig_hist_HI_out}
\end{figure*}

\subsection{\hi\ depletion time: global vs. within the stellar disc}
\label{sec:result1}

Fig.~\ref{fig_HI_dep_scaling} presents the relationships between \hi\ depletion time and several galaxy properties: stellar mass ($M_{\rm \star}$), stellar surface density ($\mu_{\rm \star}$), \nuvi\ colour, sSFR, and average \hi\ surface density ($\Sigma_{\rm HI}$).
The top two rows show the relations measured globally and within {\riso} with galaxy properties measured on the corresponding spatial scale, and their mean values (logarithms of \hi\ depletion times; Table \ref{tab_depHI_means}), which are replotted in the bottom panel for easier comparisons.
Fourteen galaxies with unreliable NUV magnitudes due to faint signal (NUV > 25 mag) are excluded for the \nuvi\ relation.
We remind readers that, for galaxies with $R_{\rm HI}<R_{\rm 25,c}$, all measurements within the stellar disc are taken within {\rhi} rather than {\riso}.
As in \citetalias{Lee2025}, we also show the relations obtained within $R_{\rm 24}$ for 617 galaxies with $R_{\rm 24}$ > 15", which is measured within the stellar isophotal radius at 24 mag arcsec$^{-2}$ in {\iband}, in the same way as for {\riso}.
Galaxies are colour-coded by the number of beams across the stellar major axis, $R_{\rm 25(24)}$/15", measured at the original resolution, from grey to darker shades, to account for possible resolution effects.
The age of the Universe of 13.8 Gyr \citep{Planck2020} and mean \htwo\ depletion time for main-sequence galaxies of 0.95 Gyr \citep{Saintonge2017} are presented as references.

For the global \hi\ depletion time (top row), the average value is 7.9 Gyr, broadly in line with previous studies with {\hi}-selected samples.
A significant fraction of WALLABY galaxies with low stellar surface density have depletion times even longer than the age of the Universe.
We find strong anti-correlations with stellar mass ($\varrho=-0.55$) and stellar surface density ($\varrho=-0.72$), i.e., galaxies with higher stellar mass and stellar surface density have shorter \hi\ depletion times.
These correlations are highly significant (p-value $\approx$ 0).
In contrast, the correlation with \nuvi\ is weak ($\varrho=-0.33$), consistent with the \nuvr\ trend for star-forming xGASS galaxies \citep{Saintonge2017}, while correlations with sSFR and $\Sigma_{\rm HI}$ are not statistically significant (p-value $\mathrm{=0.26}$ and $0.18$, respectively).
Although we might expect \hi\ depletion time to correlate with sSFR given their shared dependence on SFR, our results show no such trend, perhaps partly due to the limited dynamic range of sSFR in our sample ($\sim$1 dex).
We revisit this in Section \ref{sec:discus2}.
Marginally resolved galaxies ($R_{\rm 25}<30$") have global \hi\ depletion times that are on average 40\% longer than those of better-resolved galaxies.
This is unlikely caused by resolution effects, since the measurements use global quantities. Instead, it reflects a population difference: these galaxies tend to be {\hi}-rich galaxies at higher redshift \citepalias[see Fig.~2 in][]{Lee2025}. 
Nonetheless, the overall correlations remain consistent and not affected by spatial resolution.

When restricting to \riso\ and $R_{\rm 24}$ (second and third rows), the \hi\ depletion time shortens on average by 1.4 and 2.7 Gyr, respectively.
Correlations with stellar mass, and especially stellar surface density, become stronger ($\Delta\varrho$ = 0.05 and 0.08 for {\riso}, respectively), which is also seen in \citet{Wang2017}, while dependence on colour and sSFR remains weak.
The weak correlation with {\nuvi} diminishes further from global to within \riso\ and $R_{\rm 24}$, likely because \hi\ correlates more strongly with dust-unobscured star formation in the outer disc, where UV emission dominates \citep{Bigiel2010b}.
In contrast, a weak correlation emerges with the average \hi\ surface density within \riso\ and $R_{\rm 24}$.
However, the trend with \hi\ surface density within \riso\ disappears when controlling for stellar mass (not shown; $\varrho\sim$ 0.1), suggesting that a structural relation between \hi\ surface density and galaxy size primarily drives the trend rather than SFE.

We tested how these trends change if one uses $R_{\rm 90\%}$ (the radius enclosing 90\% of the flux in {\iband}) instead of {\riso}, since \riso\ may enclose a progressively smaller fraction of the stellar disc for galaxies with lower stellar surface brightness.
We found that the trends remain largely the same, except for the average \hi\ surface density (Fig.~\ref{fig_KS_in_r90}), which also shows only a weak correlation.

Even after excluding \hi\ beyond the stellar disc, many galaxies still show long depletion times, close to the age of the Universe.
One possibility is that some extraplanar \hi\ is still projected against the disc and therefore included in the measurement. 
However, this effect is likely small, given that \hi\ discs are typically thin and sharply truncated beyond a few scale heights \citep[scale heights $\lesssim$ 0.3--0.5 kpc;][]{Sancisi1979,Bacchini2019,Randriamampandry2021}.
Thus, these results indicate that a substantial fraction of \hi\ within the stellar disc remains in a non-star-forming phase, and that SFE on these spatial scales depends on additional factors beyond \hi\ availability.
Nonetheless, the clearer trends observed within \riso\ and $R_{\rm 24}$ emphasise the important role of \hi\ within the stellar disc in regulating star formation.
To explore the drivers of these trends, especially the role of stellar surface density, we turn to the KS relation.

\begin{figure*}
\centering
\includegraphics[width=\linewidth]{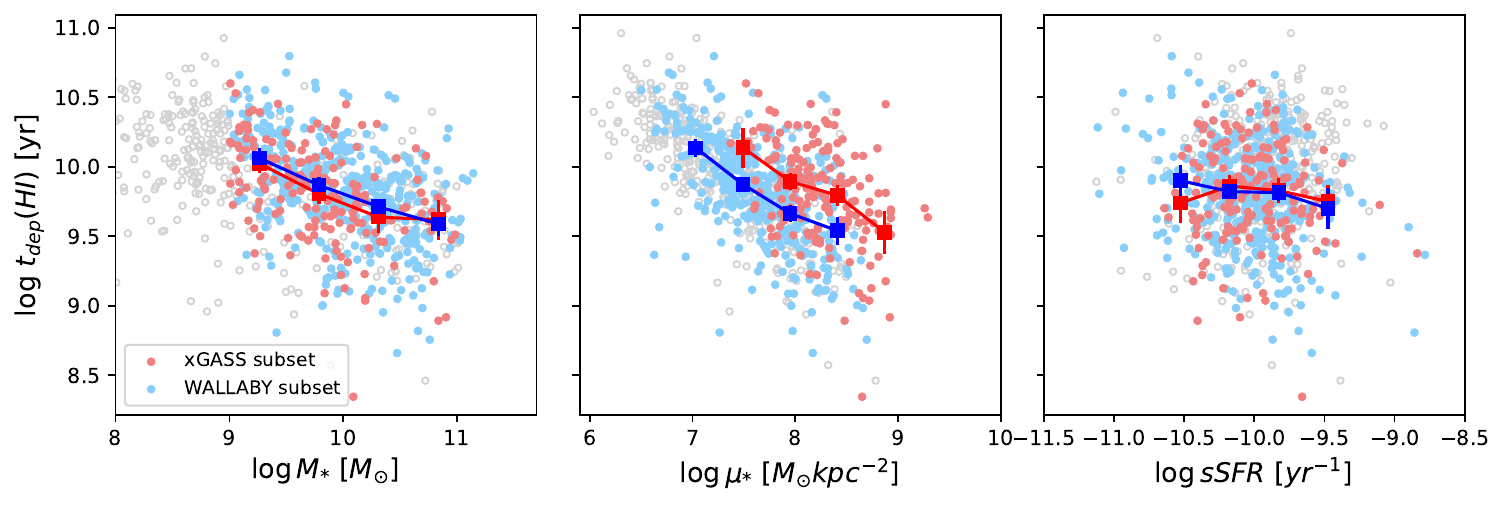}
\caption{Global \hi\ depletion time scaling relations for xGASS (red) and WALLABY (blue) galaxies, matched in \hi\ mass-redshift and stellar mass–SFR planes, limited to galaxies with stellar discs larger than the WALLABY beam. Squares show the average of $\log (t_{\rm dep})$ in each bin containing at least 10 galaxies, with error bars indicating the standard error of the mean. Grey open circles mark unmatched WALLABY galaxies, excluded from median calculation.}
\label{fig_HI_dep_x}
\end{figure*}

\subsection{Kennicutt-Schmidt relation}
\label{sec:result2}

The KS relation is the relationship between SFR and gas densities, with its slope reflecting how efficiently gas is converted into stars, corresponding to the gas depletion time (or the inverse, SFE).
This section examines the KS relation to understand how \hi\ depletion time depends on stellar surface density.

Fig.~\ref{fig_KS_in} presents the relationships between average SFR and \hi\ surface densities measured within \rhi\ (left) and \riso\ (right).
Galaxies are colour-coded by stellar surface density and grouped into four bins with similar numbers of galaxies in each bin; their means are indicated by squares to highlight how the KS relation varies with stellar surface density.
We confirmed that these results are not affected by spatial resolution: well-resolved galaxies ($R_{\rm 25}>30$") show weaker (due to limited statistics) but consistent trends.
Grey dotted lines mark depletion times of 1 and 10 Gyr, while the black dashed line shows the global KS relation from \citet{Kennicutt1998}.
Unless the stellar surface density is fixed, both KS relations within \rhi\ and {\riso} show only weak correlations between $\Sigma_{\rm SFR}$ and $\Sigma_{\rm HI}$ ($\varrho$ = 0.28 and 0.21, respectively), consistent with previous studies \citep[e.g.][]{Kennicutt1998,Bigiel2008,Schruba2011}.
At a given $\Sigma_{\rm HI}$, $\Sigma_{\rm SFR}$ spans over an order of magnitude, indicating that the presence of \hi\ alone does not determine the star formation.

When restricted to the stellar disc (right panel of Fig.~\ref{fig_KS_in}), the median relations at fixed stellar surface density become aligned with the lines of constant depletion time (grey dotted lines).
In other words, \hi\ is converted into stars with nearly constant efficiency at fixed stellar surface density.
This change of slopes arises because $\Sigma_{\rm HI}$ span a wider dynamic range within the stellar disc than in global measurements.
Removing the diffuse outer \hi\ shifts galaxies with extended \hi\ discs to higher $\Sigma_{\rm HI}$, while $\Sigma_{\rm SFR}$ change only modestly, since star formation is largely confined to the stellar disc.
Note that the main trends remain unchanged when using $R_{\rm 90\%}$ instead of {\riso}, particularly for the majority of galaxies with stellar surface density above 7~{\msunpc} (Fig.~\ref{fig_KS_in_r90}).
The KS relations binned by stellar surface density for \hi\ within \riso\ (Fig.~\ref{fig_KS_in}, right panel) and within $R_{\rm 90\%}$ (Fig.~\ref{fig_KS_in_r90}, right panel) are provided in the Appendix (Table \ref{tab_KS_means}).


As in Fig.~\ref{fig_KS_in} but in a different projection, Fig.~\ref{fig_mu_star_sfr} plots average SFR surface density within {\riso} against stellar surface density (the trend is similar for {\rhi}).
The correlation is significantly stronger ($\varrho $= 0.65) than in the KS relation, consistent with previous findings for star-forming galaxies \citep[e.g.][]{Liu2018,Lin2019,Morselli2020,Pessa2022}.
At fixed $\Sigma_{\rm HI,R25}$, the strong correlation remains, with $\Sigma_{\rm SFR,R25}$ increasing systematically.
Together, these results indicate that SFR surface density correlates more strongly with stellar surface density than with \hi\ surface density.
The mean values shown in Fig. \ref{fig_mu_star_sfr} are given in Table \ref{tab_SFR_star_means}.

The behaviour of nearly constant \hi\ depletion times at fixed stellar surface density is similar to that of molecular gas, which shows an almost uniform depletion time of $\sim$1-2 Gyr in the local Universe \citep[e.g.][]{Bigiel2008,Schruba2011,Leroy2013,Leroy2025} with only subtle variations across different galactic environments, physical scales \citep[at least as long as individual giant molecular clouds remain unresolved; e.g.][]{Bolatto2011,Schruba2011,Querejeta2021,Ellison2021a,Pessa2022} or in dwarf galaxies \citep[e.g.][]{Wyder2009,Bigiel2010b}.
This raises a question: why does {\hi}---despite not being the direct fuel for star formation---show such a strong connection to SFR through stellar surface density, mirroring the behaviour of molecular gas?
We address this in more detail in Section \ref{sec:discus2}.

\subsection{\hi\ depletion time beyond the stellar disc}
\label{sec:result3}

While Section \ref{sec:result1} and \ref{sec:result2} investigate \hi\ depletion times within the stellar disc, a substantial fraction of \hi\ lies outside this region \citepalias[about 32\% of \hi\ resides outside {\riso} with significant variation up to 80\%;][]{Lee2025}, prompting the question of how effectively the outer-disc \hi\ participates in the star formation cycle.

Fig.~\ref{fig_hist_HI_out} shows histograms of global \hi\ depletion time (left) and \hi\ depletion time within and outside \riso\ (right).
SFR measurements outside the stellar disc are highly uncertain.
Some galaxies have very low SFRs beyond \riso\ (i.e. 15 galaxies have log SFR$_{\rm out}$ [$\mathrm{M_{\rm \odot}yr^{-1}}$] < -3), and five galaxies have unmeasurable \hi\ depletion times in the outer regions, which are excluded from the calculation.
Here we focus on the relative difference between inner and outer depletion times rather than their absolute values.

We find that the median \hi\ depletion time outside the stellar disc is 15.7 Gyr, which is $\sim$9 Gyr longer than within it, with some reaching $\sim$100 Gyr, consistent with previous studies \citep{Bigiel2010b,Wang2024}.
This contrast highlights the dependence of \hi\ depletion time on location within the galaxy.
The global \hi\ depletion times average over both dense, star-forming regions and diffuse outer discs, and thus depletion times confined within the stellar disc provide a more physically meaningful measurement.

\section{Discussion}
\label{sec:discus}

\subsection{Global \hi\ depletion time scaling relations in the literature}
\label{sec:discus1}

Our analysis of 841 WALLABY galaxies shows that \hi\ depletion times correlate most strongly with stellar surface density (and secondly, stellar mass), while correlations with \nuvi\ colour, sSFR, and average \hi\ surface density are weak or absent (Fig.~\ref{fig_HI_dep_scaling}).

To compare with xGASS \citep{Catinella2018}, we overlay xGASS galaxies on our global \hi\ depletion time scaling relations in Fig.~\ref{fig_HI_dep_x}.
For a fair comparison, we match the WALLABY and xGASS galaxies by restricting both samples to similar regions of the \hi\ mass-redshift and SFR-stellar mass planes, resulting in the xGASS subsample lying on the star-forming main sequence.
We also apply a similar sample selection criterion to the xGASS subsample ($R_{\rm 90\%}$ > 15").
The relations with stellar mass and sSFR agree well, while the stellar surface density shows a systematic offset of $\sim$0.5 dex along the x-axis largely due to intrinsic differences between the xGASS and WALLABY selected galaxies.
In particular, at higher redshift ($z>0.025$), xGASS mainly includes massive galaxies ($\rm > 10^{10}M_{\odot}$), while WALLABY spans a broader stellar-mass range from $\rm \sim 10^{9}M_{\odot}$ to $\rm 10^{11}M_{\odot}$.
Differences in effective radii due to different photometry and survey depth might further contribute to the discrepancy in stellar surface density.

We note that several selection effects may influence these trends.
The exclusion of marginally resolved or unresolved \hi\ detections may bias both samples against compact, {\hi}-poor systems that could have shorter depletion times, particularly at low stellar masses.
Additionally, our sample is representative mainly below $\rm M_{\rm \star}\sim10^{10}M_{\odot}$ (Fig.~\ref{fig_sfms}), while more massive and less star-forming galaxies are underrepresented.
Including them would likely steepen the observed trend with stellar mass and stellar surface density.
Despite these biases, the correlation with stellar surface density, even when passive systems are included by using survival analysis \citep[Fig.~6 in][]{Saintonge2022}, suggests that the trend is physical rather than purely selection-driven.

These results also align with previous {\hi}-selected surveys \citep[e.g.][]{Huang2012,Jaskot2015,Wong2016,Wang2017,Tudorache2024}, although their strength may vary depending on the sample size.
For example, ALFALFA galaxies in \citet{Huang2012} showed weak positive correlations between {\hi}-based SFE and stellar mass.
\citet{Jaskot2015} reported a strong dependence on stellar surface density, interpreted as evidence that mid-plane pressure regulates the atomic-to-molecular gas conversion efficiency and thus the efficiency of star formation \citep{Blitz2006}, with its secondary dependence on sSFR at fixed stellar mass.

\begin{figure}
\centering
\includegraphics[width=0.9\linewidth]{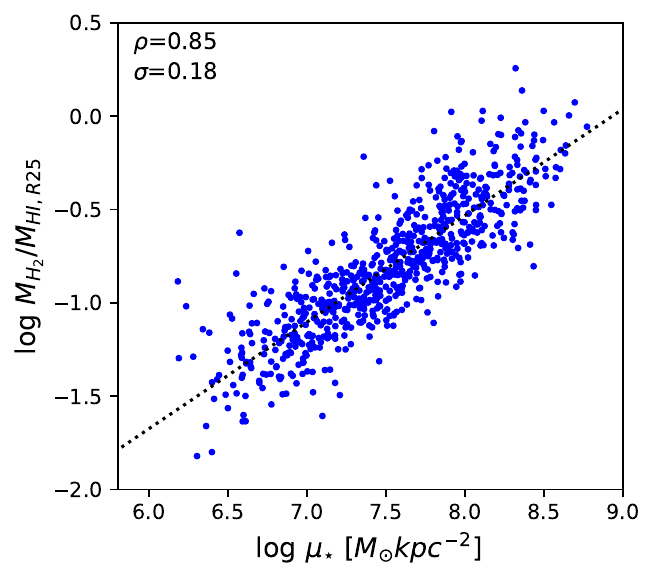}
\caption{Relation between (estimated) {\htwo}-to-{\hi} mass ratios and stellar surface density, using \hi\ within {\riso}. The black dotted line is the linear regression fit, with the Pearson coefficient and scatter (i.e. standard deviation along the y-axis from the fitted line) indicated in the upper left corner.}
\label{fig_Rmol}
\end{figure}

The agreement with previous studies and theoretical expectations suggests that stellar surface density may play an important role in setting the \hi\ depletion time.
Nonetheless, a homogeneous analysis across both {\hi}- and stellar-mass–selected samples will be important to fully quantify the impact of sample selection on the derived scaling relations.


\subsection{Linking \hi\ within the stellar disc to molecular gas}
\label{sec:discus2}

\hi\ is not the direct fuel for star formation.
However, its depletion time, which traces how long the \hi\ reservoir would last at the current SFR, remains nearly constant at fixed stellar surface density (Fig.~\ref{fig_KS_in}), similar to that of molecular gas.
To investigate the link between atomic and molecular gas, we estimate molecular gas masses by using the empirical median relation between molecular gas fraction and sSFR for main-sequence galaxies from xCOLD GASS \citep[$\varrho$ = 0.80;][]{Saintonge2017} for 703 WALLABY galaxies with $-10.9 < $log sSFR [\Msun\ yr$^{-1}$]$ < -9.66$ (range of the adopted relation).

Fig.~\ref{fig_Rmol} shows how the estimated {\htwo}-to-{\hi} mass ratio (within the stellar disc) varies with stellar surface density for the sub-sample.
Galaxies with higher stellar surface density have higher {\htwo}-to-{\hi} mass ratio ($\varrho$ = 0.85), reflecting a more efficient conversion of atomic gas into molecular gas.
The scatter increases slightly when global \hi\ masses are used ($\Delta\sigma$ = 0.05; not shown), likely because most molecular gas is located within the star-forming disc and thus \hi\ and \htwo\ are more co-spatial within the stellar disc.
When the {\hi}-based KS relation within {\riso} (right panel in Fig.~\ref{fig_KS_in}) is converted to a molecular gas-based relation, it yields an almost constant molecular gas depletion time of $\sim$1 Gyr (Fig.~\ref{fig_KS_mol}).
We note that these results are largely a consequence of the adopted sSFR–molecular gas fraction relation.
Nevertheless, they support the interpretation that the behaviour of \hi\ at fixed stellar surface density is mainly driven by variations in the efficiency of atomic–to–molecular gas conversion, combined with the nearly constant molecular gas depletion time of $\sim$1 Gyr.

\begin{figure}
\centering
\includegraphics[width=\linewidth]{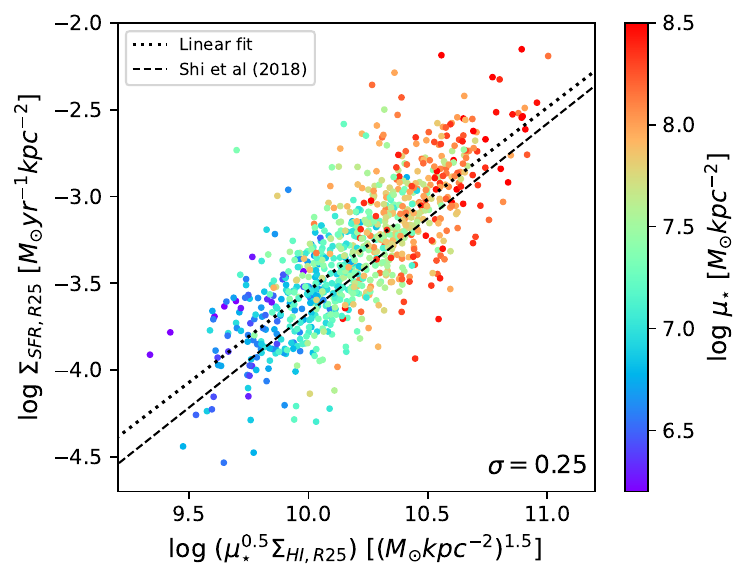}
\caption{Extended Kennicutt-Schmidt relation, with average SFR and \hi\ surface densities measured within {\riso} (or \rhi\ if $R_{\rm HI}<R_{\rm 25,c}$). Galaxies are colour-coded by their stellar surface density. The dotted line shows the linear regression fit, and the scatter is shown in the bottom right corner. The dashed line presents the best-fit relation based on the total gas surface density from \citet{Shi2018}.}
\label{fig_exKS_in}
\end{figure}

This positive correlation between {\htwo}-to-{\hi} mass ratio and stellar surface density has been observed from global \citep{Catinella2018} to kpc scales \citep{Leroy2008,Wong2013,Eibensteiner2024}, although this trend may not hold for low-stellar-mass galaxies \citep{Wong2016}.
Stellar surface density serves as an effective tracer of regions where higher densities correspond to denser, more shielded gas and enhanced \htwo\ formation.
Physically, this can be explained by the mid-plane pressure model suggested by \citet{Elmegreen1989}, where local stellar and gas gravity increases hydrostatic pressure, promoting {\hi}--to--{\htwo} conversion \citep[e.g.][]{Wong2002,Blitz2004,Blitz2006,Leroy2008,Ostriker2010,Sun2020}.
In addition, dust shielding in dense stellar environments attenuates dissociating UV radiation, further supporting a higher molecular gas fraction \citep[e.g.][]{Krumholz2009,Krumholz2013}.

Highlighting the important role of the stellar component in regulating star formation through its contribution to the mid-plane pressure, \citet{Shi2011,Shi2018} proposed the extended KS relation, in which the SFR surface density is plotted against the total gas (atomic and molecular gas) surface density multiplied by the square root of stellar surface density.
They demonstrated that this relation holds remarkably well across a broad range of environments (e.g. outer discs of dwarfs, local spirals, giant molecular clouds) on a sub-kiloparsec scale, and even for integrated measurements of high-redshift star-forming and starburst galaxies.

Fig.~\ref{fig_exKS_in} shows the extended KS relation for our WALLABY sample, with average SFR and \hi\ surface densities within {\riso}.
The separations between galaxies with different stellar surface densities, seen as nearly parallel sequences in the standard KS relation (Fig.~\ref{fig_KS_in}), collapse into a single linear trend in the extended KS relation.
This is driven by the strong correlation between SFR and stellar surface densities ($\sigma=0.29$; Fig.~\ref{fig_mu_star_sfr}).
Our fitted slope closely matches the best-fit relation based on the total gas surface density from \citet{Shi2018}, with a slight offset ($\sim0.1$ dex).
The inclusion of molecular gas (\htwo\ and helium) can largely explain this difference.
We interpret this relation as strong evidence that the mid-plane pressure proxy on the x-axis effectively traces the molecular gas surface density, thereby producing a molecular-gas-like KS relation even when only \hi\ is measured.
Incorporating molecular gas surface densities may further strengthen this correlation.

\begin{figure}
\centering
\includegraphics[width=\linewidth]{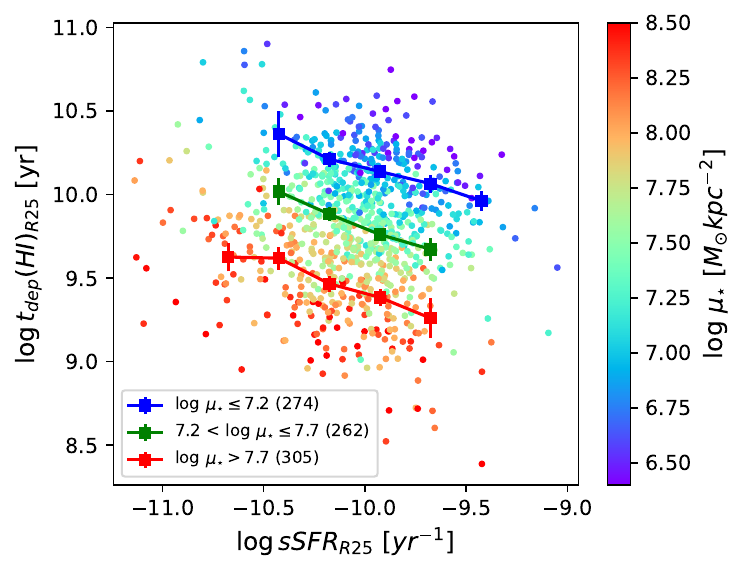}
\caption{Relation between \hi\ depletion time and sSFR within {\riso} (or \rhi\ if $R_{\rm HI}<R_{\rm 25,c}$) colour-coded by stellar surface density. Squares show the average of $\log (t_{\rm dep})$ in each stellar surface density bin, with the bin range and number of galaxies indicated in the lower left corner. Error bars represent the standard error of the mean.}
\label{fig_HI_dep_ssfr}
\end{figure}

Stellar surface density therefore provides a strong physical link between {\hi}, {\htwo}, and star formation. 
Given that both \hi\ depletion time and sSFR depend on the SFR, one might also expect an anti-correlation between \hi\ depletion time and sSFR.
However, this trend appears weak in our scaling relations in Fig.~\ref{fig_HI_dep_scaling}.
Fig.~\ref{fig_HI_dep_ssfr} helps clarify this behaviour by showing the relation between \hi\ depletion time and sSFR, colour-coded by stellar surface density, with means in three stellar surface density bins; the values are listed in Table \ref{tab_depHI_sSFR_means}.
At fixed stellar surface density, a negative correlation emerges ($\varrho < -0.4$).
This indicates that stellar surface density is the primary driver of \hi\ depletion time, likely by tracing regions of efficient {\hi}--to--{\htwo} conversion, where star formation is enhanced.
\hi\ within such regions (at least inside the stellar disc) serves as a key intermediate, linking the extended atomic reservoir to the dense molecular gas that fuels star formation.

\subsection{Star formation beyond the stellar disc}
\label{sec:discus3}

Deep observations reveal star formation extending beyond the stellar disc in the form of extended UV (and sometimes H$\alpha$) discs \citep[e.g.][]{Thilker2005,Thilker2007,Gil_de_Paz2005,Gil_de_Paz2007,Bigiel2010a}.
However, the conversion of \hi\ into \htwo\ in these regions is highly inefficient, resulting in \hi\ depletion times roughly 9 Gyr longer than in the inner regions and in some cases reaching $\sim$100 Gyr \citep[Fig.~\ref{fig_hist_HI_out}; see also][]{Bigiel2010b,Wang2024}.
These timescales by far exceed the Hubble time, even though molecular gas depletion times remain at $\sim$1-2 Gyr \citep{Schruba2011}.
This low efficiency is expected given the physical conditions in outer discs: the weak stellar potential makes it difficult for gas to collapse; low metallicity and dust content reduce the shielding needed for \htwo\ formation \citep[][and references therein]{Schinnerer2024}; gas-disc flaring spreads gas diffusely over large vertical scales \citep[e.g.][]{Vollmer2016,Mancera_Pina2022}, and environmental processes further hinder collapse \citep[e.g.][]{Cortese2021}.
Under these conditions, \hi\ remains largely atomic, and star formation proceeds only slowly and stochastically.

In addition, environmental processes influence the \hi\ content of galaxies, particularly in the outer regions of the disc.
Although WALLABY's sensitivity limits detections of galaxies that have undergone severe gas stripping \citep{Reynolds2022}, some galaxies in our sample likely have experienced mild environmental effects.
We find that WALLABY galaxies with truncated \hi\ discs ($R_{\rm HI}<R_{\rm 25,c}$; 109 galaxies) generally have shorter \hi\ depletion times within \rhi\ (mean $t_{\rm dep}(\rm HI)$ = 3 Gyr), primarily due to lower \hi\ masses while maintaining similar SFRs.
This is consistent with the findings of \citet[][their Fig.~13]{Cortese2021}, and suggests that environmental effects can enhance the apparent SFE by removing diffuse outer {\hi} while leaving the inner star-forming gas largely intact, yielding similar scaling relations but systematically lower \hi\ content.
The improved statistics offered by the full WALLABY survey will allow us to explore these trends in more detail.

The outer \hi\ reservoir, while inefficient at forming stars in situ, may nevertheless serve as a long-term fuel source through processes such as accretion.
Moreover, studies imply that atomic gas may play a more direct role in regulating star formation in this regime, where it is often the dominant baryonic component: \citet{Bigiel2010b} found a stronger {\hi}--SFR correlation in outer than in inner discs, resembling that observed in dwarf galaxies, while \citet{Wang2024} proposed a link between {\hi}-based SFE and stellar surface density in outer discs. 
We find a similar trend in our sample, such as a stronger correlation between \hi\ depletion time and \hi\ surface density and a tighter {\hi}-based KS relation outside the stellar disc, but the uncertainties (primarily in the SFR) are too large to draw firm conclusions, so we do not present these results here.
A comprehensive exploration of the role of outer-disc gas in fueling star formation will be enabled by the higher-resolution \hi\ data and large statistics expected from the future Square Kilometre Array (SKA) observations.

\section{Conclusions}
\label{sec:conclus}

In this work, we examined how \hi\ depletion time and its scaling relations change when \hi\ is restricted to the stellar disc ({\riso}) by measuring \hi\ mass, stellar mass and SFR within matched physical scales for 841 galaxies from the WALLABY pilot survey.
We investigated the Kennicutt-Schmidt (KS) relation to further understand the regulation of \hi\ and star formation within the stellar disc, depending on the physical conditions of galaxies.
Our main findings are:

\begin{itemize}
    \item On average, the global \hi\ depletion time of 7.9 Gyr shortens by 1.4 Gyr within the stellar disc, yet many galaxies still show depletion times longer than the Hubble time, implying that a substantial fraction of the \hi\ remains in a non-star-forming phase even in star-forming regions. 
    \item We find that \hi\ depletion times anti-correlate strongly with stellar surface density, and this correlation becomes even tighter when restricted to the stellar disc, indicating a closer connection between \hi\ within the stellar disc and star formation than for the global \hi\ reservoir.
    \item The KS relations show that, \textit{at fixed stellar surface density,} \hi\ depletion time is nearly constant within the stellar disc, likely because the efficiency of converting \hi\ into \htwo\ remains roughly fixed under similar conditions, combined with the near-universal molecular gas depletion time. This underscores stellar surface density as a good tracer of the conditions under which \hi\ is efficiently converted into molecular gas and ultimately into stars.
    \item Beyond the stellar disc, \hi\ depletion times are on average $\sim$10 Gyr longer than within it, reflecting the inefficient star formation in outer, low-density, low-metallicity environments. Global \hi\ depletion times average over very different regimes, whereas measurements restricted to the stellar disc provide a more physically meaningful view of the link between \hi\ and star formation.
\end{itemize}

Taken together, our results suggest that the ability of \hi\ to act as a fuel for star formation depends critically on its spatial location and the local physical conditions. 
Separating "active" \hi\ in the disc from outer reservoirs, where conversion of \hi\ into stars is less efficient and the gas is more affected by environment, may have broader cosmological implications, improving scaling relations, clustering analysis, and interpretations of \hi\ intensity mapping observations \citep[e.g.][]{Kovetz2017,Villaescusa_Navarro2018}.
The full WALLABY survey will extend this analysis to a larger number of galaxies, but the combination of large statistics, sensitivity and resolution needed to move beyond the simple distinction of \hi\ within and beyond \riso\ adopted here will need the full SKA.

\section*{Acknowledgements}
We thank the anonymous referee for constructive comments that improved the paper.
This scientific work uses data obtained from Inyarrimanha Ilgari Bundara, the CSIRO Murchison Radio-astronomy Observatory. We acknowledge the Wajarri Yamaji People as the Traditional Owners and native title holders of the Observatory site. CSIRO’s ASKAP radio telescope is part of the Australia Telescope National Facility (https://ror.org/05qajvd42). Operation of ASKAP is funded by the Australian Government with support from the National Collaborative Research Infrastructure Strategy. ASKAP uses the resources of the Pawsey Supercomputing Research Centre. Establishment of ASKAP, Inyarrimanha Ilgari Bundara, the CSIRO Murchison Radio-astronomy Observatory and the Pawsey Supercomputing Research Centre are initiatives of the Australian Government, with support from the Government of Western Australia and the Science and Industry Endowment Fund.

WALLABY acknowledges technical support from the Australian SKA Regional Centre (AusSRC).

Parts of this research were supported by the Australian Research Council Centre of Excellence for All Sky Astrophysics in 3 Dimensions (ASTRO 3D), through project number CE170100013.

LC acknowledges support from the Australian Research Council Discovery Project funding scheme (DP210100337).

\section*{Data Availability}

The WALLABY source catalogue and associated data products (e.g. cubelets, moment maps, integrated spectra, radial surface density profiles) are available online through the CSIRO ASKAP Science Data Archive (CASDA) and the Canadian Astronomy Data Centre (CADC). All source and kinematic model data products are mirrored at both locations. Links to the data access services and the software tools used to produce the data products as well as documented instructions and example scripts for accessing the data are available from the WALLABY Data Portal (\url{https://wallaby-survey.org/data/}).
 



\bibliographystyle{mnras}
\bibliography{reference} 




\appendix

\section{Supplementary materials}
\label{sec_appendix}

\subsection{Comparison between scaling relations within \riso\ and $R_{\rm 90\%}$}

To test the impact of different definitions of the stellar disc, we repeat our analysis using $R_{\rm 90\%}$ instead of {\riso}, as \riso\ may enclose a smaller fraction of the stellar disc for galaxies with lower stellar surface brightness.
We use the subset of 761 galaxies with both $R_{\rm 25}>15$" and $R_{\rm 90\%}>15$".

\begin{figure*}
\centering
\includegraphics[width=0.85\linewidth]{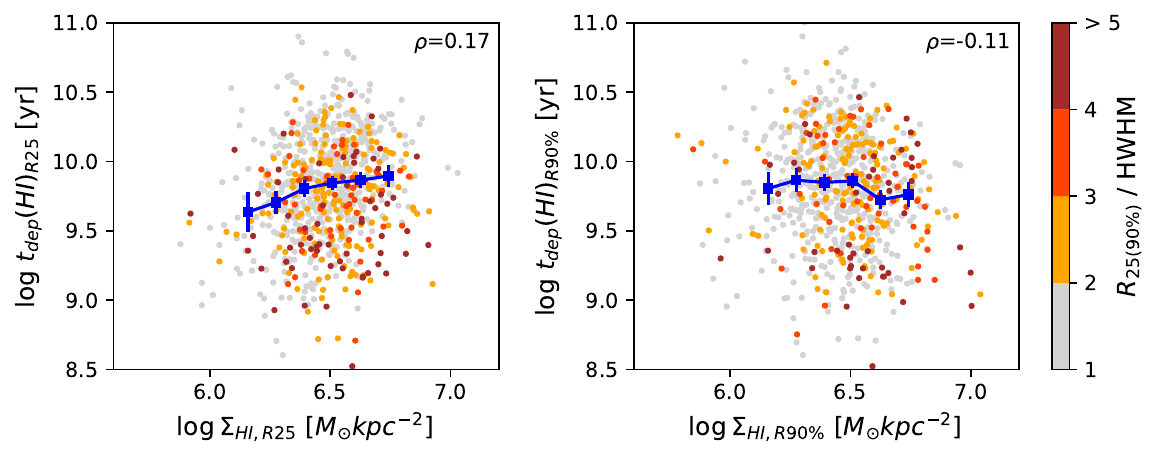}
\caption{Scaling relations of \hi\ depletion time measured within \riso\ (left) and $R_{\rm 90\%}$ (right) as a function of average \hi\ surface density at the corresponding spatial scale. Symbols and colours are the same as in Fig.~\ref{fig_HI_dep_scaling}.}
\label{fig_HI_dep_scaling_r90}
\end{figure*}

Overall, the \hi\ depletion time scaling relations (Fig.~\ref{fig_HI_dep_scaling}) remain largely unchanged, except for differences in the average \hi\ surface density.
Fig.~\ref{fig_HI_dep_scaling_r90} compares the average \hi\ surface density relations, measured within \riso\ (left) and $R_{\rm 90\%}$ (right).
The weak positive correlation observed within \riso\ becomes slightly negative within $R_{\rm 90\%}$, but both remain very weak ($|\varrho|<0.2$).
Marginally resolved galaxies ($R_{\rm 90\%}$ < 30"; grey points) tend to have lower average \hi\ surface densities within $R_{\rm 90\%}$ compared to better resolved galaxies ($R_{\rm 90\%}$ > 30"; coloured points), suggesting that beam smearing affects the measurement based on $R_{\rm 90\%}$ more systematically.

\begin{figure*}
\centering
\includegraphics[width=0.85\linewidth]{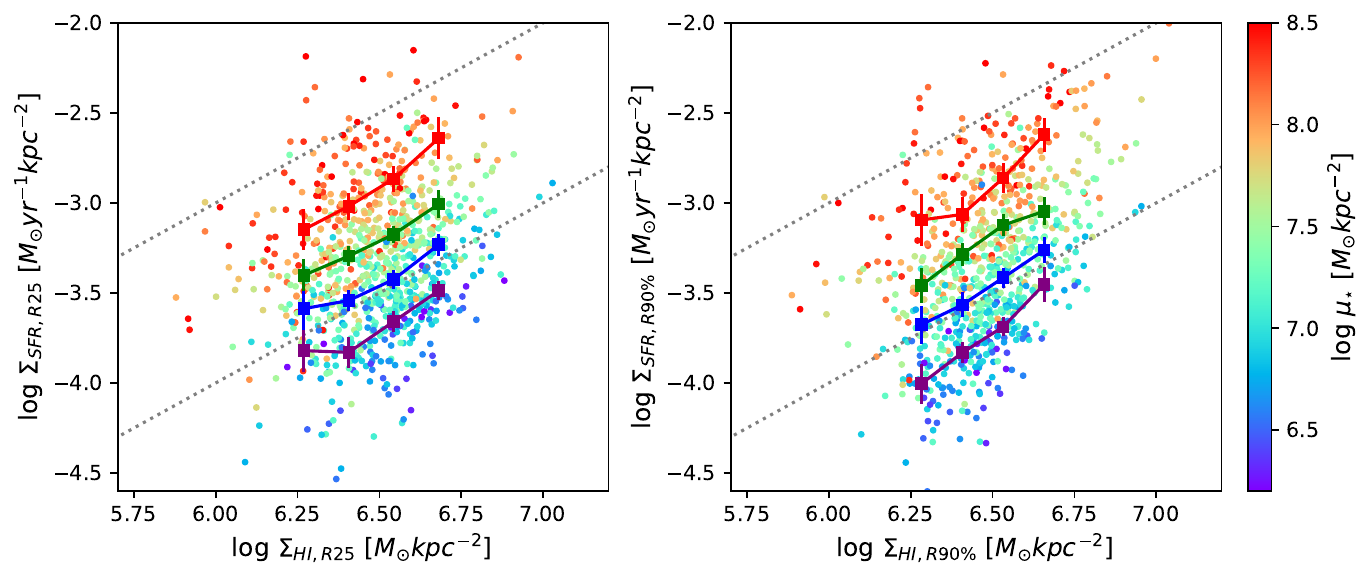}
\caption{Scaling relations between average SFR and \hi\ surface densities measured within \riso\ (left) and $R_{\rm 90\%}$ (right). Symbols and colours are the same as in Fig.~\ref{fig_KS_in}.}
\label{fig_KS_in_r90}
\end{figure*}

Fig.~\ref{fig_KS_in_r90} presents the KS relations using average SFR and \hi\ surface densities within \riso\ (left) and $R_{\rm 90\%}$ (right), corresponding to Fig.~\ref{fig_KS_in} in the main text.
Using $R_{\rm 90\%}$ shifts galaxies with low stellar surface density (< 7 {\msunpc}) toward lower surface densities, as these galaxies generally have larger $R_{\rm 90\%}$/\riso\ ratios.
Importantly, the main trends remain unchanged, particularly for the majority of galaxies with stellar surface densities above 7 {\msunpc}.

\subsection{Molecular gas Kennicutt-Schmidt relation}

\begin{figure}
\centering
\includegraphics[width=0.85\linewidth]{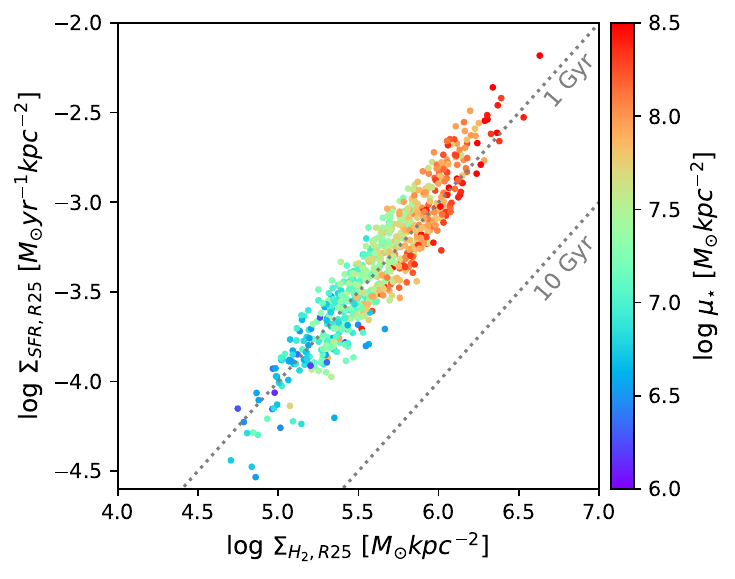}
\caption{Scaling relation between average SFR and (estimated) H$_2$ surface densities, measured within {\riso}. Galaxies are colour-coded by their stellar surface density. The grey dotted diagonal lines indicate \hi\ depletion times of 1 and 10 Gyr.}
\label{fig_KS_mol}
\end{figure}

Fig.~\ref{fig_KS_mol} shows the relation between average SFR and \textit{estimated {\htwo}} surface densities for WALLABY galaxies.
As discussed in the main text, the inferred \htwo\ masses follow the adopted molecular gas fraction--sSFR relation \citep{Saintonge2017}.
Therefore, this figure does not independently recover the molecular gas KS relation, but is included to illustrate the expected behaviour of molecular gas and reinforce the interpretation presented in Section \ref{sec:discus2}.

\subsection{Scaling relations tables}

Tables presenting the mean scaling relations shown in the main text are provided below.

\begin{table*}
\caption{Scaling relations of \hi\ depletion time measured globally, within {\riso}, and within $R_{\rm 24}$ (Fig.~\ref{fig_HI_dep_scaling}). Columns: (1) bin centre; (2), (4), (6) number of galaxies; (3), (5), (7) mean log $t_{\rm dep}$({\hi}) and standard error of the mean.}
\begin{threeparttable}
\centering
\begin{tabular}{|cc|cc|cc|cc|}
\hline
 & & \multicolumn{2}{c}{Global} & \multicolumn{2}{c}{\riso} & \multicolumn{2}{c}{$R_{\rm 24}$}\\
 \cmidrule(lr){3-4}\cmidrule(lr){5-6}\cmidrule(lr){7-8}
 & $x$ & N & <log $t_{\rm dep}$({\hi})> & N & <log $t_{\rm dep}$({\hi})> & N & <log $t_{\rm dep}$({\hi})> \\
 & & & $[\rm yr]$ & & $[\rm yr]$ & & $[\rm yr]$ \\
 & (1) & (2) & (3) & (4) & (5) & (6) & (7)\\
\hline
 $\log M_{\rm \star}$ & 7.98 & 43 & 10.20 $\pm$ 0.05 & 45 & 10.11 $\pm$ 0.05 & 28 & 9.98 $\pm$ 0.05\\
 $[\rm M_{\odot}]$    & 8.55 & 108 & 10.12 $\pm$ 0.03 & 115 & 10.07 $\pm$ 0.03 & 82 & 10.00 $\pm$ 0.03\\
                      & 9.12 & 162 & 10.06 $\pm$ 0.02 & 162 & 9.96 $\pm$ 0.02 & 110 & 9.86 $\pm$ 0.03\\
                      & 9.68 & 186 & 9.86 $\pm$ 0.03 & 183 & 9.78 $\pm$ 0.03 & 132 & 9.70 $\pm$ 0.03\\
                      & 10.25 & 205 & 9.72 $\pm$ 0.02 & 201 & 9.61 $\pm$ 0.02 & 161 & 9.54 $\pm$ 0.02\\
                      & 10.82 & 119 & 9.61 $\pm$ 0.03 & 112 & 9.49 $\pm$ 0.03 & 90 & 9.45 $\pm$ 0.03\\
\hline
 $\log\mu_{\rm \star}$ & 6.57 & 66 & 10.31 $\pm$ 0.03 & 66 & 10.26 $\pm$ 0.03 & 34 & 10.25 $\pm$ 0.04\\ 
 $[$\msunkpc$]$        & 6.90 & 131 & 10.19 $\pm$ 0.02 & 131 & 10.11 $\pm$ 0.02 & 85 & 10.07 $\pm$ 0.02\\
                       & 7.23 & 162 & 10.01 $\pm$ 0.02 & 162 & 9.95 $\pm$ 0.02 & 108 & 9.91 $\pm$ 0.03\\
                       & 7.57 & 178 & 9.84 $\pm$ 0.02 & 178 & 9.76 $\pm$ 0.02 & 147 & 9.70 $\pm$ 0.02\\
                       & 7.90 & 158 & 9.67 $\pm$ 0.02 & 158 & 9.56 $\pm$ 0.02 & 128 & 9.50 $\pm$ 0.02\\
                       & 8.23 & 98 & 9.53 $\pm$ 0.04 & 98 & 9.40 $\pm$ 0.03 & 81 & 9.34 $\pm$ 0.03\\
\hline
 \nuvi\              & 1.65 & 89 & 10.08 $\pm$ 0.03 & 57 & 9.88 $\pm$ 0.03 & 20 & 9.72 $\pm$ 0.07\\ 
 $\mathrm{[mag]}$    & 2.15 & 223 & 9.99 $\pm$ 0.02 & 182 & 9.92 $\pm$ 0.03 & 70 & 9.75 $\pm$ 0.03\\
                     & 2.65 & 206 & 9.88 $\pm$ 0.03 & 254 & 9.84 $\pm$ 0.02 & 182 & 9.77 $\pm$ 0.03\\
                     & 3.15 & 155 & 9.76 $\pm$ 0.03 & 175 & 9.72 $\pm$ 0.03 & 150 & 9.70 $\pm$ 0.03\\
                     & 3.65 & 67 & 9.66 $\pm$ 0.05 & 92 & 9.65 $\pm$ 0.04 & 105 & 9.64 $\pm$ 0.04\\
                     & 4.15 & 38 & 9.76 $\pm$ 0.04 & 41 & 9.63 $\pm$ 0.05 & 47 & 9.60 $\pm$ 0.06\\
\hline
 $\log$ sSFR          & -10.68 & 35 & 9.94 $\pm$ 0.07 & 37 & 9.80 $\pm$ 0.07 & 48 & 9.82 $\pm$ 0.06\\ 
 $[\rm yr^{-1}]$      & -10.45 & 55 & 9.91 $\pm$ 0.05 & 79 & 9.87 $\pm$ 0.04 & 81 & 9.81 $\pm$ 0.04\\
                      & -10.22 & 177 & 9.86 $\pm$ 0.03 & 210 & 9.80 $\pm$ 0.03 & 185 & 9.73 $\pm$ 0.03\\
                      & -9.98 & 261 & 9.88 $\pm$ 0.02 & 284 & 9.81 $\pm$ 0.02 & 195 & 9.69 $\pm$ 0.03\\
                      & -9.75 & 213 & 9.90 $\pm$ 0.03 & 158 & 9.76 $\pm$ 0.03 & 68 & 9.57 $\pm$ 0.05\\
                      & -9.52 & 67 & 9.94 $\pm$ 0.04 & 47 & 9.81 $\pm$ 0.05 & 21 & 9.63 $\pm$ 0.08\\
\hline
 $\log\Sigma_{\rm HI}$ & 6.16 & 128 & 9.88 $\pm$ 0.03 & 40 & 9.55 $\pm$ 0.07 & 20 & 9.48 $\pm$ 0.11\\ 
 $[$\msunkpc$]$        & 6.28 & 230 & 9.93 $\pm$ 0.03 & 115 & 9.68 $\pm$ 0.04 & 46 & 9.48 $\pm$ 0.06\\
                       & 6.39 & 290 & 9.92 $\pm$ 0.02 & 169 & 9.80 $\pm$ 0.03 & 100 & 9.67 $\pm$ 0.04\\
                       & 6.51 & 129 & 9.80 $\pm$ 0.03 & 241 & 9.84 $\pm$ 0.02 & 117 & 9.73 $\pm$ 0.04\\
                       & 6.62 & 23 & 9.71 $\pm$ 0.07 & 180 & 9.90 $\pm$ 0.03 & 189 & 9.77 $\pm$ 0.03\\
                       & 6.74 & -- & -- & 55 & 9.89 $\pm$ 0.04 & 95 & 9.82 $\pm$ 0.04\\
\hline
\end{tabular}
\label{tab_depHI_means}
\end{threeparttable}
\end{table*}

\begin{table*}
\caption{Scaling relations between average \hi\ and SFR surface densities measured within $R_{\rm 25}$ (Fig.~\ref{fig_KS_in}, right panel) and $R_{\rm 90\%}$ (Fig.~\ref{fig_KS_in_r90}, right panel). Columns: (1), (4) number of galaxies; (2), (5) bin centre; (3), (6) mean log($\Sigma_{\rm SFR}$) and standard error of the mean.}
\begin{threeparttable}
\centering
\begin{tabular}{ c c c c c c c }
\hline
 & \multicolumn{3}{c}{\riso} & \multicolumn{3}{c}{$R_{\rm 90\%}$}\\
 \cmidrule(lr){2-4}\cmidrule(lr){5-7}
 & N & $\log\Sigma_{\rm HI}$ & <$\log\Sigma_{\rm SFR}$> & N & $\log\Sigma_{\rm HI}$ & <$\log\Sigma_{\rm SFR}$> \\
 & & [\msunkpc]  & [\msunyrkpc] & & [\msunkpc] & [\msunyrkpc] \\
& (1) & (2) & (3) & (4) & (5) & (6)\\
\hline
 $\log\mu_{\rm \star}\leq7$ & 17 & 6.27 & -3.83 $\pm$ 0.06 & 31 & 6.28 & -4.00 $\pm$ 0.06\\ 
 $[$\msunkpc$]$                           & 36 & 6.41 & -3.82 $\pm$ 0.04 & 56 & 6.41 & -3.83 $\pm$ 0.03\\
                            & 64 & 6.54 & -3.66 $\pm$ 0.03 & 64 & 6.53 & -3.69 $\pm$ 0.03\\
                            & 50 & 6.68 & -3.49 $\pm$ 0.03 & 18 & 6.66 & -3.45 $\pm$ 0.05\\
\hline
 $7<\log\mu_{\rm \star}\leq7.5$ & 28 & 6.27 & -3.57 $\pm$ 0.05 & 33 & 6.28 & -3.68 $\pm$ 0.05\\ 
 $[$\msunkpc$]$                           & 63 & 6.41 & -3.56 $\pm$ 0.03 & 66 & 6.41 & -3.57 $\pm$ 0.03\\
                            & 93 & 6.54 & -3.43 $\pm$ 0.03 & 72 & 6.53 & -3.42 $\pm$ 0.03\\
                            & 37 & 6.68 & -3.23 $\pm$ 0.03 & 30 & 6.66 & -3.26 $\pm$ 0.04\\
\hline
 $7.5<\log\mu_{\rm \star}\leq7.9$ & 34 & 6.27 & -3.39 $\pm$ 0.04 & 24 & 6.28 & -3.46 $\pm$ 0.05\\ 
 $[$\msunkpc$]$                           & 57 & 6.41 & -3.26 $\pm$ 0.03 & 51 & 6.53 & -3.29 $\pm$ 0.03\\
                            & 65 & 6.54 & -3.17 $\pm$ 0.03 & 54 & 6.53 & -3.12 $\pm$ 0.03\\
                            & 33 & 6.68 & -3.03 $\pm$ 0.04 & 25 & 6.66 & -3.05 $\pm$ 0.04\\
\hline
 $\log\mu_{\rm \star}>7.9$ & 50 & 6.27 & -3.11 $\pm$ 0.05 & 30 & 6.28 & -3.09 $\pm$ 0.07\\ 
 $[$\msunkpc$]$                           & 58 & 6.41 & -3.04 $\pm$ 0.04 & 41 & 6.53 & -3.06 $\pm$ 0.05\\
                            & 51 & 6.54 & -2.88 $\pm$ 0.04 & 53 & 6.53 & -2.86 $\pm$ 0.04\\
                            & 15 & 6.68 & -2.64 $\pm$ 0.06 & 22 & 6.66 & -2.62 $\pm$ 0.05\\
\hline
\end{tabular}
\label{tab_KS_means}
\end{threeparttable}
\end{table*}

\begin{table}
\caption{Relation between average SFR surface density within {\riso} and stellar surface density (Fig.~\ref{fig_mu_star_sfr}). Columns: (1) number of galaxies; (2) bin centre; (3) mean log($\Sigma_{\rm SFR,R25}$) and standard error of the mean.}
\begin{threeparttable}
\centering
\begin{tabular}{ c c c c }
\hline
 & N & $\log\mu_{\rm \star}$ & <$\log\Sigma_{\rm SFR,R25}$> \\
 & & [\msunkpc]  & [\msunyrkpc]\\
 & (1) & (2) & (3) \\
\hline
 $\log\Sigma_{\rm HI,R25}\leq6.35$ & 17 & 6.65 & -3.93 $\pm$ 0.07\\ 
 $[$\msunkpc$]$                           & 31 & 7.15 & -3.60 $\pm$ 0.05\\
                            & 66 & 7.65 & -3.44 $\pm$ 0.03\\
                            & 67 & 8.15 & -3.18 $\pm$ 0.04\\
\hline
 $6.35<\log\Sigma_{\rm HI,R25}\leq6.5$ & 28 & 6.65 & -3.83 $\pm$ 0.05\\ 
 $[$\msunkpc$]$                           & 63 & 7.15 & -3.58 $\pm$ 0.03\\
                            & 93 & 7.65 & -3.25 $\pm$ 0.03\\
                            & 37 & 8.15 & -3.01 $\pm$ 0.04\\
\hline
 $6.5<\log\Sigma_{\rm HI,R25}\leq6.6$ & 34 & 6.65 & -3.72 $\pm$ 0.05\\ 
 $[$\msunkpc$]$                          & 57 & 7.15 & -3.48 $\pm$ 0.03\\
                            & 65 & 7.65 & -3.21 $\pm$ 0.03\\
                            & 33 & 8.15 & -2.92 $\pm$ 0.04\\
\hline
 $\log\Sigma_{\rm HI,R25}>6.6$ & 50 & 6.65 & -3.48 $\pm$ 0.03\\ 
 $[$\msunkpc$]$                           & 58 & 7.15 & -3.25 $\pm$ 0.03\\
                            & 51 & 7.65 & -3.03 $\pm$ 0.03\\
                            & 15 & 8.15 & -2.71 $\pm$ 0.07\\
\hline
\end{tabular}
\label{tab_SFR_star_means}
\end{threeparttable}
\end{table}

\begin{table}
\caption{Relation between \hi\ depletion time and sSFR within {\riso} (Fig.~\ref{fig_HI_dep_ssfr}). Columns: (1) number of galaxies; (2) bin centre; (3) mean $\log t_{\rm dep}$({\hi})$_{\rm R_{25}}$ and standard error of the mean.}
\begin{threeparttable}
\centering
\begin{tabular}{ c c c c }
\hline
 & N & $\log \rm sSFR_{R25}$ & <$\log t_{\rm dep}$({\hi})$_{\rm R_{25}}$>\\
 & & [$\rm yr^{-1}$]  & [yr]\\
 & (1) & (2) & (3)\\
\hline
 $\log\mu_{\rm HI,R25}\leq7.2$ & 15 & -10.42 & 10.36 $\pm$ 0.07\\ 
 $[$\msunkpc$]$               & 60 & -10.18 & 10.21 $\pm$ 0.02\\
                            & 112 & -9.93 & 10.14 $\pm$ 0.02\\
                            & 59 & -9.68 & 10.06 $\pm$ 0.03\\
                            & 19 & -9.43 & 9.96 $\pm$ 0.03\\
\hline
 $7.2<\log\mu_{\rm HI,R25}\leq7.7$ & 28 & -10.42 & 10.02 $\pm$ 0.04\\ 
 $[$\msunkpc$]$                           & 93 & -10.18 & 9.88 $\pm$ 0.02\\
                            & 92 & -9.93 & 9.76 $\pm$ 0.02\\
                            & 34 & -9.68 & 9.67 $\pm$ 0.04\\
\hline
 $\log\mu_{\rm HI,R25}>7.7$ & 29 & -10.68 & 9.63 $\pm$ 0.04\\ 
 $[$\msunkpc$]$                           & 54 & -10.42 & 9.62 $\pm$ 0.03\\
                            & 93 & -10.18 & 9.47 $\pm$ 0.02\\
                            & 83 & -9.93 & 9.38 $\pm$ 0.03\\
                            & 29 & -9.68 & 9.26 $\pm$ 0.06\\
\hline
\end{tabular}
\label{tab_depHI_sSFR_means}
\end{threeparttable}
\end{table}

\bsp	
\label{lastpage}
\end{document}